\documentclass[12pt]{article}
\ifx\pdfoutput\undefined
\usepackage[dvips,bookmarks]{hyperref}
\else
\usepackage{hyperref}
\fi
\hypersetup{colorlinks,bookmarksopen,bookmarksnumbered,citecolor=blue,
    pdfstartview=FitH}
\def\hhref#1{\href{http://arxiv.org/abs/hep-th/#1}{hep-th/#1}}
\def\hhpref#1{\href{http://arxiv.org/abs/hep-ph/#1}{hep-ph/#1}}
\def\mhref#1{\href{mailto:#1}{#1}}
\usepackage{amssymb,amsfonts,amsmath,color,graphicx}

\parskip=\medskipamount            
\def\mysection#1{\section{#1}
    \setcounter{equation}{0}
    \renewcommand{\theequation}{\thesection.\arabic{equation}}}
\def\mysubsection#1{\subsection{#1}
    \setcounter{equation}{0}
    \renewcommand{\theequation}{\thesubsection.\arabic{equation}}}

\def\on#1#2{{\buildrel{\mkern2.5mu#1\mkern-2.5mu}\over{#2}}}
\def\f#1#2{{\textstyle{#1\over#2}}}    
\def\slap#1#2{\setbox0=\hbox{$#1{#2}$}
    #2\kern-\wd0{\hbox to\wd0{\hfil$#1{/}$\hfil}}}
\def\sla#1{\mathpalette\slap{#1}}                

\def\normalord#1{\mathopen{:}#1\mathclose{:}}

\begin{document}

\begin{center}

{March 28, 2006 \hfill YITP-SB-06-5} \vskip.5in
{\bf{\huge\color{cyan}Simpler superstring scattering}}\\[.3in]
Kiyoung Lee and Warren Siegel\footnote{ \it
\mhref{klee@insti.physics.sunysb.edu} {\rm and}
\mhref{siegel@insti.physics.sunysb.edu}}\\[.1in]
{\it C. N. Yang Institute for Theoretical Physics \\
State University of New York, Stony Brook, NY 11794-3840}\\[.5in]

\end{center}

\begin{abstract}

We give a new, manifestly spacetime-supersymmetric method for
calculating superstring scattering amplitudes, using the ghost
pyramid, that is simpler than all other known methods.  No
pictures nor non-vertex insertions are required other than the
usual $b$ and $c$ ghosts of the bosonic string.  We evaluate some
tree and loop amplitudes as examples.

\end{abstract}

\thispagestyle{empty}           
\newpage

\mysection{Introduction}

Many formalisms have been introduced for calculating scattering
amplitudes for superstrings.  The most practical of these have
been (covariant) Ramond-Neveu-Schwarz (RNS) \cite{rns},
(lightcone) Green-Schwarz (GS) \cite{gs}, hybrid RNS-GS (H)
\cite{h}, and pure spinor (PS) \cite{ps}.  All of these have (at
least) two important defects:

\noindent (1) Some kind of insertion is required.  It may be
separate from the vertices, or may be combined with some vertices
to put them into different ``pictures".  The result is to
complicate the calculations or destroy manifest symmetry.  (The
only exception is tree graphs with external bosons only, where
such methods make cyclic symmetry more obscure but avoid producing
extra terms that cancel.)

\noindent (2) Supersymmetry is not completely manifest.  The most
serious case is RNS, where fermion vertices are much more
complicated than boson (because the spinors are not free fields,
so in practice noncovariant exponentials of bosons must be used),
and sums over spin structures (periodic/antiperiodic boundary
conditions) must be performed in loops.  In the GS and H cases
there is partial supersymmetry (and partial 10D Lorentz
invariance), which complicates vertices for the ``longitudinal"
directions, which are required for general higher-point
calculations; for this reason we will not consider GS and H in
detail.  The most symmetric is PS, which has only an integration
measure that is explicitly dependent on the spinor coordinates.

In a previous paper \cite{us} we introduced a new formalism for
the superstring (based on a similar one for the superparticle
\cite{ws}) using an infinite pyramid of ghosts for the spinor
coordinate (GP) \cite{oldgp}.  A derivation was also given from a
covariant action.  (The RNS action is not spacetime-supersymmetry
covariant.  The GS action \cite{gsaction} has defied covariant
quantization \cite{badaction}.  The H and PS formalisms do not
follow from the quantization of an action with general worldsheet
metric.)  The Becchi-Rouet-Stora-Tyutin operator found there was
rather complicated, but fortunately none of the results of our
previous paper will be needed explicitly here for calculation, but
only for justification of the validity of our approach.  In fact,
the gauge-fixed action and massless vertex operators were guessed
much earlier \cite{current}.  (An early attempt to apply them to
amplitude calculations failed because spinor ghosts were not
included \cite{badguess}.)  The fact that these simple rules can
be applied so naively hints that perhaps an even simpler formalism
exists that implies the same rules.

There are (at least) two new conceptual results in this paper (in
addition to the explicit calculations), both of which involve the
treatment of zero-modes.  These allow us to evaluate trees and
loops without evaluating explicit integrals or (super)traces over
these zero-modes, thereby solving the above two problems:

\noindent (1) In loop calculations we infrared regularize the
worldsheet propagators.  In principle one should do this anyway,
since IR divergences are notorious in two dimensions, especially
for 2D conformal field theories, but usually such problems are
avoided by examining only IR-safe quantities.  In our case such a
regularization allows a simple counting of the infinite number of
zero-modes arising from the ghost pyramid (including those from
the physical spinor), with the only result being the introduction
of factors of 1/4 due to the usual summation $1-2+3-...=1/4$.
(Regularization of $x$ zero-modes is unnecessary; it only replaces
the momentum-conservation $\delta$-function with a sharp
Gaussian.)

\noindent (2) In tree graphs these zero-modes do not appear
separately, having been absorbed into the definition of the
(first-quantized) vacuum.  Specifically, since we do not perform
explicit integration over spinor zero-modes, we also do not need
to define measure factors for such integrations, make insertions
of operators (essentially Dirac $\delta$-functions in those modes)
to kill those modes, nor use operators of different pictures to
hide such insertions.  We do not make special manipulations to
deal with such modes; care of them is taken automatically by
naively ignoring them.  Although we do not analyze this vacuum (or
other) state in detail here (we effectively work with the old
Heisbenberg matrix mechanics, ignoring Schr\"odinger wave
functions), we explain why such behavior is implied by the
standard N=1 superspace formulation of the vector multiplet.

The net result of these ideas is that the calculational rules are
the most naive generalization of the rules of the bosonic string:
(1) The $b$ and $c$ ghosts appear in the same way, affecting only
the measure.  (2) The spinor ghosts serve only to ensure correct
counting of zero-modes, and give an extra factor of 1/4 to any
trace of $\gamma$-matrices.  (3) IR regularization takes care of
all (physical and ghost) spinor zero-modes.  (4) The vertex
operator for the massless states generalizes the bosonic-string
one just by adding the same spin terms as in ordinary field theory
or supergraphs (to include the spinor vertices), taking into
account the stringy generalization of the algebra of covariant
derivatives \cite{current}.

Consequently, for the case of tree graphs with external vectors
only, our rules are almost identical to (R)NS calculations in the
${\cal F}_1$ picture.  We explain the advantage of this picture
and why it is more relevant to the superstring.

As an interesting side result, we show how the $\partial\theta$ terms in the $DP\Omega$ current algebra arise already in the superparticle.

\mysection{Rules}

\mysubsection{Vertex operators}

We now present the main result of this paper, the rules
themselves, with examples later.  (Derivations are given in the
Appendices.)  Here we will calculate amplitudes with only massless
external states.  (We also concentrate on open strings, but the
results generalize in the usual way to closed.)

To a limited extent first-quantization can be applied to particles
as well as to strings:  It gives only one-particle irreducible
graphs (vertices at the tree level), whereas for the string it
gives complete S-matrix amplitudes by duality (for given loop
level and external states).  However, the methods are almost
identical, particularly since the superparticle is the zero-modes
of the superstring.

The vertex operators follow from the results of our previous paper
\cite{us} but are basically those of \cite{current} with a small
modification from ghosts (as expected from the integrated vertex
operators of PS \cite{ps}):
$$ V = A^A (x,\theta) J_A $$ where
$A^A$ are superfields and $J_A$ are 2D currents:
$$ A^A = (A_\alpha, A^a, W^\alpha, F^{ab}) $$
$$ J_A = (\Omega^\alpha, P_a, D_\alpha, \hat S_{ab}) $$
where $J_A$ have zero-modes $j_A$, of which only $p_a$ and
$d_\alpha$ act nontrivially on $A^A$.  $D,P,\Omega$ are the
currents of \cite{current}, while $\hat S$ is the Lorentz current
of the $\theta$ ghosts (``superspin"). (Appendix \ref{hl} gives
the relation of vertex operators between Lagrangian and
Hamiltonian formalisms.)

As for the bosonic string, the integrated vertex operator is $\int
V$ and the unintegrated one is $cV$; the $b$ and $c$ ghosts work
in exactly the same way, to keep the measure conformal.  (We could
also add a term $\alpha'(\partial_a A^a)\partial c$ to the
unintegrated vertex operator to avoid having to apply $\partial_a
A^a=0$ \cite{haidong}.)

The external-state superfields and the currents can be expanded in
$\theta$ for evaluation in terms of 2D Green functions of the
fundamental variables:  For example, the vertex for just the
vector is then
$$ V_B = A_a(x)\partial x^a +\f12 F^{ab}(x) S_{ba} $$
where $S$ is the Lorentz current of all $\theta$'s, physical and
ghost.  There are also terms higher-order in $\theta$, but in the
absence of external fermions there are no $\pi$'s to cancel the
extra $\theta$'s, so such terms won't contribute. Because of its
universality, this form is useful for comparison to other
formalisms.

\mysubsection{Current algebra}

However, when calculating general amplitudes (including fermions),
it is more convenient to expand neither the currents nor
superfields (thus manifesting supersymmetry). This requires rules
for evaluating products of arbitrary numbers of currents.
Although this problem is generally intractable for arbitrary
representations of arbitrary current algebras, in our case it is
relatively simple:

\noindent (1) $\hat S$ doesn't act on the superfields.  It is
quadratic in free fields, so the matrix element of any product of
such currents is simply the sum of products of loops of them (in
2D perturbation theory), from contracting the (ghost) $\theta$ of
one with the $\pi$ of the next.  Each such loop contributes the
trace of the product of the $\gamma$ matrices that appear
sandwiched between $\theta$ and $\pi$ in $\hat
S_{ab}=\theta\gamma_{ab}\pi|_>$ (where ``$|_>$" means to restrict
to ghosts).

\noindent (2) The remaining currents $D_\alpha$, $P_a$, and
$\Omega^\alpha$ form a separate algebra.  Although their ``loops"
are more complicated (since $D$ is cubic in free fields), the
structure constants are so simple that no loop contains more than
4 currents: only the combinations $PP$, $D\Omega$, $DDP$, or
$DDDD$.  Since $D$ and $P$ (but not $\Omega$) can also act on
superfields, the matrix element of such currents and superfields
reduces to the sum of products of these 4 types with strings of
$D$ and $P$ acting on superfields.

The loops are:
\begin{eqnarray}
\label{simplerule}
\langle P_{a}(1)P_{b}(2)\rangle&=&-\eta_{ab}G_{x}''(1,2)\nonumber\\
\langle D_{\alpha}(1)\Omega^{\beta}(2)\rangle&=&-i\delta_{\alpha}^{\beta}G_{\theta}'(1,2)\nonumber\\
\langle D_{\alpha}(1)D_{\beta}(2)P_{a}(3)\rangle&=&-i\gamma_{a\alpha\beta}[2G_{\theta}(2,3)G_{\theta}'(1,3)-2G_{\theta}(1,3)G_{\theta}'(3,2)\nonumber\\
&&+G_{\theta}(1,2)(G_{x}''(1,3)+G_{x}''(2,3))]\nonumber\\
\langle D_{\alpha}(1)D_{\beta}(2)D_{\gamma}(3)D_{\delta}(4)\rangle&=&2iG_{\theta}'(1,2)G_{\theta}(1,3)G_{\theta}(1,4)(\gamma^{a}_{\alpha\gamma}\gamma_{a\delta\beta}-\gamma^{a}_{\alpha\delta}\gamma_{\gamma\beta})\nonumber\\
&&+2iG_{\theta}'(1,2)G_{\theta}(2,3)G_{\theta}(2,4)(\gamma^{a}_{\beta\delta}\gamma_{\gamma\alpha}-\gamma^{a}_{\beta\gamma}\gamma_{a\delta\alpha})\nonumber\\
&&+iG_{x}''(1,2)\left[G_{\theta}(1,3)G_{\theta}(2,4)\gamma^{a}_{\alpha\gamma}\gamma_{a\beta\delta}\right.\nonumber\\
&&\left.~~-G_{\theta}(2,3)G_{\theta}(1,4)\gamma^{a}_{\beta\gamma}\gamma_{a\alpha\delta}\right]+\mbox{perm.}
\end{eqnarray}
where $\langle\quad\rangle$ refers to fully contracted operator
products, and  ``1" means ``$z_{1}$", etc. We have distinguished
the $x$ and $\theta\pi$ Green functions ($G_{x}$ and $G_{\theta}$)
because only $G_{\theta}$ gives zero-mode corrections, which is
explained in detail in Appendix \ref{ir}. For N string loops, $G$
is a genus-N Green function: for trees,
$G_{x}'(z_{1}-z_{2})=-iG_{\theta}(z_{1}-z_{2})=-\frac{1}{z_{1}-z_{2}}$
and
$G_{x}''(z_{1}-z_{2})=-iG_{\theta}'(z_{1}-z_{2})=\frac{1}{(z_{1}-z_{2})^{2}}$;
at 1 string loop they are Jacobi theta functions and their
derivatives; etc.

The action of the currents on the fields is given by considering
all possible symmetrizations of the $D$'s.  Any symmetrization of
2 $D$'s (acting on a field) gives
\begin{equation}
D_{(\alpha}(1)D_{\beta)}(2)\quad\to\quad G_{\theta}(1,2)
\gamma^a_{\alpha\beta}[P_a(1) +P_a(2)]
\end{equation}
This reduces any
string of currents to sums of strings of $P$'s times
antisymmetrized strings of $D$'s, which are evaluated as
\begin{equation}
D_{[\alpha}(1)\cdots D_{\beta]}(2)P_a(3)\cdots P_b(4)A(5) =
G_{\theta}(1,5)\cdots G_{x}'(4,5) (d_{[\alpha}\cdots
d_{\beta]}p_a\cdots p_b A)(5)
\end{equation}
where $p_a=-i\partial_a$, and we can replace
$\pi_\alpha=\partial/\partial\theta^\alpha$ with the usual
supersymmetry covariant derivative $d_\alpha$ in such
antisymmetrizations since final results can always be evaluated at
$\theta=0$ by supersymmetry.

By 10D dimensional analysis, any $\hat S$ loop is dimensionless,
while any $DP\Omega$ loop has dimension 2.  This implies (contrary
to expectations, but well known from the bosonic case) that each
$DP\Omega$ loop carries an extra factor of {\it the inverse of}
$\alpha'$.  (In the particle case, there is instead an inverse of
$z$.)  Thus, the maximum number of $DP\Omega$ loops gives the
lowest power in momenta, and each loop less gives two more powers
of momenta.  (One way to see the dimensional analysis is to note
that each current acting on a superfield gives a $G_{x}'$ or
$G_{\theta}$. The same is true in a $DP\Omega$ loop, except that 2
currents ``close" the loop to give a $G_{x}''$ or $G_{\theta}'$.
Thus, each $DP\Omega$ loop introduces an extra factor of
$G_{x}''(\mbox{or}~G_{\theta}')/(G_{\theta})^2$. On the other
hand, closing an $\hat S$ loop gives a $(G_{\theta})^2$ instead of
$G_{\theta}'$, so such loops give no extra factor.)

Finally, there is the usual momentum dependence coming from Green
functions connecting the superfields to each other, from their $x$
dependence only:  For the usual plane waves,
\begin{equation}
\langle A(1) \cdots A(N) \rangle = A \cdots A \ e^{-\sum_{i<j}
k_i\cdot k_j G_{x}(i,j)}
\end{equation}
with units $\alpha'=1/2$ for the string.

\mysubsection{Component expansion}

The final result for an amplitude is given as a ``kinematic
factor" times a scalar function of momentum invariants, expressed
as an integral over the worldsheet positions of the vertices.  The
kinematic factor is expressed, by the above procedure, as a sum of
products of superfields, representing external state wave
functions.  The string rules have effectively already performed
covariant $\theta$ integration, so these superfields may be
evaluated at $\theta=0$.  (As in the usual superspace methods,
where $\theta$ expansion and integration is replaced by the action
on the ``Lagrangian" of the product of all supersymmetry covariant
derivatives $d_\alpha$, supersymmetry guarantees that all $\theta$
dependence cancels, up to total $x$ derivatives.)

The evaluation of spinor derivatives follows from the (linearized)
constraints on the gauge covariant superspace derivatives, and
their Bianchi identities \cite{higher}.  The result is
\begin{eqnarray}
d_{(\alpha}A_{\beta)} & = & 2\gamma^a_{\alpha\beta}A_a \nonumber\\
d_\alpha A_a - \partial_a A_\alpha & = & 2\gamma_{a\alpha\beta}W^\beta \nonumber\\
d_\alpha W^\beta & = & \f12\gamma^{ab}{}_\alpha{}^\beta F_{ba}  \nonumber\\
d_\alpha F^{ab} & = &
2i\gamma^{[a}_{\alpha\beta}\partial^{b]}W^\beta
\end{eqnarray}
The result is also (linearized) gauge invariant (except for
$\partial_a A^a=0$, as explained above), so one may use a
Wess-Zumino gauge where $A_\alpha=0$ at $\theta=0$. (A review of
gauge covariant derivatives appears in Appendix \ref{ce}.)

\mysubsection{IR regularization}

In evaluation of tree graphs there is the usual $\delta$-function
for conservation of total momentum from the zero-modes of $x$, but
$\theta$ effectively has no zero-modes: The effect of the $\theta$
ghosts is to mimic GS where, unlike momentum, the 8 surviving
fermionic variables of the lightcone are self-conjugate, and thus
have no vanishing eigenvalues.  Thus there is no residual
integration over $\theta$ zero-modes (unlike PS).

In loops there is the usual summation over $\theta$ zero-modes in
the sum over all states, but the ghosts again mimic GS by
effectively reducing the number to 8 from the physical 32
($\theta$ and conjugate $\pi$), using the sum
$$ 1 - 2 + 3 - 4 +
... = 1/4 $$ when counting the number of $\theta$'s at successive
ghost levels (alternating in statistics).  Application of this
rule requires infrared regularization of the 2D Green functions to
``remove" the zero-modes:  The factor in the partition function
from these zero-modes is the IR regulator $\epsilon$ to the power
$16\times 1/4 = 4$ (from the 16-valued spinor index on the
$\theta$'s).  Since the $\theta\pi$ Green function goes as
$1/\epsilon$ (+ the usual finite expression + ${\cal
O}(\epsilon)$), the amplitude vanishes until 4-point.  Thus the
power of the regulator counts zero-modes.

The 1/4 rule also applies in $\gamma$-matrix algebra.  Amplitudes
involve traces of products of $\gamma$-matrices.  These matrices
are the same at each ghost level (except that chirality, as well
as statistics, alternates with ghost level), so the net effect of
the ghosts appears only when taking a trace:  Applying the usual
$\gamma$-matrix identities, the trace is reduced to $str(I) =
16\times 1/4 = 4$, again reproducing GS.  The difference from GS
is that the $\gamma$-matrices are for 10 dimensions, so the result
is Lorentz covariant, and the usual 10D Levi-Civita tensor is
produced (where appropriate) instead of spurious 8D
$\epsilon$-tensors.  For example, anomalies can be found from
6-point graphs. (Details of the regularized Green functions are
given in Appendix \ref{ir}.)

\mysection{Trees}

\mysubsection{RNS pictures}

We begin by proving that the trees with external bosons are
identical to those obtained from (the NS sector of) RNS.  This is
most obvious in the ${\cal F}_1$ picture.  Although this picture
was the original one to be used in (R)NS amplitude calculations,
it was immediately replaced with the ${\cal F}_2$ picture
\cite{nst}.  We refer here to the picture for the physical
coordinates ($x,\psi$), and not just the ghosts:  For example,
vector vertices have always been $\partial x+...$ except for two
$\psi$ vertices, while in the ${\cal F}_1$ picture all vertices
are $\partial x+...$\nobreak\ .  In the proof of equivalence
\cite{nst}, starting from the ${\cal F}_1$ picture, one pulls
factors of (the $\pm 1/2$ modes of) $G=\psi\cdot\partial x$
(worldsheet supersymmetry generator) off of two unintegrated
$\partial x+...$ vertices to turn them into $\psi$ vertices, then
collides the $G$'s to produce (the 0 mode of) a worldsheet
energy-momentum tensor $T$, which gives a constant acting on a
physical state.  (With ghosts the approach is similar, with $G$
replaced with the picture-changing operator, which is simply the
operator product of the gauge-fixed $G$ with $e^{\phi}$ in terms
of the bosonized ghost $\phi$.)  The resulting rules are then the
same as the rules for the bosonic string, including the factors of
$c$ for the three unintegrated vertices, except that the $\partial
x$ vertex has the extra spin term.  The $\beta$ and $\gamma$
ghosts are completely ignored; the vacuum used is in what is
usually called the ``$-1$ picture", so the zero-modes of $\gamma$
(or $\phi$) are already eliminated.  (What is usually called
``picture changing" in the modern covariant formalism would start
with the ${\cal F}_2$ picture, introduce two factors of picture
changing times inverse picture changing, use the picture changing
to change the two $\psi$ vertices, and use the inverses to change
the initial and final vacuua.  Unfortunately, the inverse has an
overall factor of $c$, so in the new vacuum $\langle\gamma\gamma
c\rangle\sim 1$ \cite{big}, and the $\gamma$'s pick out the $\psi$
terms again in two of the unintegrated vertices
$\gamma\psi+c(\partial x+...)$.  Thus such transformations
preserve the ${\cal F}_2$ picture as far as the physical sector is
concerned.)

Historically, the ${\cal F}_1$ picture was introduced first
because: (1) It is more similar to the bosonic string, and (2)
cyclic symmetry is manifest (no need to bother with picture
changing).  The ${\cal F}_2$ picture was then chosen because the
physical-state conditions were more obvious.  Although in modern
language the BRST conditions are clear in either picture, it's
interesting to examine the differences in the pictures if the
ghosts are ignored, since the ghosts differ in different
formulations of the superstring, but all formulations have similar
integrated vertices.  Then the ground state of the ${\cal F}_2$
picture is the ``physical" tachyon, at $m^2=-1/2$, while in the
${\cal F}_1$ picture it's an ``unphysical" tachyon at $m^2=-1$.
Furthermore, the ${\cal F}_1$ picture has an additional
``ancestor" trajectory 1/2 unit higher than the leading physical
trajectory.  These ``disadvantages" were noticed in the days
before Gliozzi-Scherk-Olive projection.  On the other hand, for
the superstring this projection eliminates the ``physical" tachyon
as well as the ancestor trajectory.  So the only remaining
additional unphysical state of the ${\cal F}_1$ picture is its
vacuum, while GSO projection has eliminated the vacuum of the
${\cal F}_2$ picture altogether!   This suggests that any
comparison of the RNS formulation to others would be easier in the
${\cal F}_1$ picture.

\mysubsection{For bosons only}

The proof of equivalence of the ${\cal F}_1$ vector trees to the
vector trees of our formalism is then simple:  One only has to
note that the operator algebra of the vertices is identical.  But
the vertices are identical in form; only the explicit
representation of the spin current is different.  So one only has
to check the equivalence of the two current algebras.  Since they
are both (10D) Lorentz currents, quadratic in free fields, this
means just checking that the central charge is the same.  (The
same method has been used for comparing PS to the ${\cal F}_2$
picture \cite{brenno}.)  The reason the result for the central
charge is the same is that the GP result is the same as the GS
result:  The $\gamma$-matrix algebra is the same except for a
trace, which is 1/4 as big in the lightcone as for a covariant
spinor, but GP again gets a factor of 1/4 from summing over
ghosts.  (As we'll see below, similar arguments apply in loops,
unless one gathers enough spin currents to produce a Levi-Civita
tensor.)

The calculations in the ${\cal F}_1$ picture (and GP) are somewhat
harder than the ${\cal F}_2$ picture because two vertices have
been replaced with ones that generate more terms, which cancel.
Also, RNS bosonic trees are simpler than PS or GP because
integration over the vector fermion $\psi^a$ effectively does all
$\gamma$-matrix algebra.  However, tree amplitudes with fermions
are much harder in RNS than PS or GP (and increase in difficulty
as the number of fermions increases).  PS is still simpler than
GP, because $\theta$ integration takes the place of the change in
the two vertices, and so also avoids generating extra terms.  So,
for trees RNS is the easiest for pure bosons, PS is easiest with
fermions, and GP is a bit harder than both.  However, GP requires
fewer rules, since all vertices are the same, so it produces more
terms at an intermediate stage but is easier to ``program".  This
feature is a peculiarity of tree graphs:  At the loop level we'll
see that GP maintains the simplest rules, while RNS produces extra
terms that cancel (because supersymmetry is not manifest).

At first sight these rules for GP might seem peculiar because
there is no explicit integral over spinor zero-modes, as expected
in known superspace approaches.  The answer can be seen from
examining the simpler (and better understood) case of 4D N=1 super
Yang-Mills.  Since the vacuum of the open bosonic string can be
identified with a constant Yang-Mills ghost (or gauge parameter),
we examine the ghost superfield $\phi$, and look at $\phi=1$, a
supersymmetric condition.  Since this superfield is chiral, and
supergraphs prefer unconstrained superfields, we write $\phi=\bar
d^2\chi$ in terms of a general complex superfield $\chi$.  Then
clearly $\chi=\bar\theta^2$ in our case.  This is still
supersymmetric because of the gauge invariance $\delta\chi=\bar
d_{\dot\alpha}\lambda^{\dot\alpha}$.  Furthermore, this $\chi$ has
a nice norm, $\int d^4\theta\ |\chi|^2=1$.  In Hilbert-space
notation we thus write the norm and supersymmetry as
$$ \langle 0 | 0 \rangle = 1 , \quad\quad
q_\alpha | 0 \rangle = Q | \lambda \rangle_\alpha $$ so the vacuum
is supersymmetry invariant up to a BRST triviality, and the norm
includes zero-mode integration, but the extra zero-modes are
absorbed by the vacuua, and no insertions are required. (We could
also use $| \lambda \rangle_\alpha=\Lambda_\alpha| 0 \rangle$ to
define $\hat q_\alpha = q_\alpha -[Q,\Lambda_\alpha]$, $\hat
q_\alpha| 0 \rangle = 0$.) This supersymmetry of the vacuum is
enough to ensure the amplitudes transform correctly, since the
vertex operators are superfields times supersymmetry invariant
currents, and the vacuum and vertex operators (integrated and
unintegrated) are BRST invariant.  (The unintegrated vertex
operators we have used are BRST invariant only after including
terms higher-order in ghost $\theta$'s, which don't contribute to
amplitudes for massless external states, and probably not for
massive ones either, because of the absence of ghost $\pi$'s to
cancel them.)  The fact that the vacuum is ``half-way" up in the
$\theta$ expansion was also found for the expansion in the spinor
ghost coordinates in a lightcone analysis of the BRST cohomology
for the GP superparticle \cite{ws}.  Note that this choice of vacuum is relevant only for trees; at 1 loop one effectively does a (super)trace over all states rather than a vacuum expectation value, so the vacuum is irrelevant.  (We assume a similar situation will occur at higher loops, but we have not checked yet.)  As we will see below, one important affect on this vacuum choice for trees, which does not affect loops, is that:

{\it For tree graphs only, background fields are always evaluated in the Wess-Zumino gauge.}

If this vacuum structure can
be better understood, it might be possible to find an analog of
the ${\cal F}_2$ picture for GP, avoiding the production of extra
canceling terms, making it the simplest formalism even for trees.
As an attempt at formulating such a picture, one can consider this picture for RNS:  For pure bosons, two vertices must be in the ``$-1$ picture", so it is convenient to consider those two as the initial and final states, using the vertex operators on the initial and final vacuua.  Generalizing only those 2 states to include fermions, we can write their vertex operators as in GP, but now identifying the currents with
$$ D_\alpha = e^{-\phi/2}S_\alpha, \quad
P_a = e^{-\phi}\psi_a, \quad \Omega^\alpha = e^{-3\phi/2}S^\alpha $$
The first two are the usual for the spinor (in the $-1/2$ picture) and vector (in the $-1$ picture), while the last can be identified as that for the spinor in the $-3/2$ picture (also with conformal weight 1) if we use the ``supersymmetric gauge"
$$ W^\alpha \sim \gamma^{a\alpha\beta}\partial_a A_\beta $$
instead of the WZ gauge.  (Hitting $cA_\alpha\Omega^\alpha$ with picture changing produces $cW^\alpha D_\alpha$.)  These currents satisfy almost the same algebra as the usual ones (including $DP\sim\Omega$, to leading order); the only exceptions are $\Omega P$ and $\Omega\Omega$.  As a guess for the GP analog, we can then try to construct a new $DP\Omega$ for this picture that depends only on the ghosts.  Unfortunately (the simplest guess for) this construction seems not to work, apparently because the dependence on the WZ gauge hasn't been eliminated, and is incompatible with the supersymmetric gauge.

\mysubsection{General 3-point}

As explained in Section 2, we prefer the superfield formalism for
the calculation of amplitudes with fermions.   This includes the
all-vector amplitude in the same calculation.  The only
nonvanishing operator products for the 3-point tree, after
applying the Landau gauge condition ($\partial\cdot A=0$), are:
\begin{eqnarray}
\label{supertree} &&A.\quad\langle P_{a}(1)P_{b}(2)\rangle \times
P_{c}(3)A^{a}(1)A^{b}(2)A^{c}(3)+
\mbox{permutations}\nonumber\\
&&B.\quad(P_{c}(3)A^{a}(1))(P_{a}(1)A^{b}(2))(P_{b}(2)A^{c}(3))+\mbox{perm.}\nonumber\\
&&C.\quad\langle P_{a}(1)P_{b}(2)\rangle \times D_{\alpha}(3)A^{a}(1)A^{b}(2)W^{\alpha}(3)+\mbox{perm.}\nonumber\\
&&D.\quad\langle D_{\alpha}(1)D_{\beta}(2)P_{a}(3)\rangle \times W^{\alpha}(1)W^{\beta}(2)A^{a}(3)+\mbox{perm.}\nonumber\\
&&E.\quad
\langle D_{\beta}(2)\Omega^{\gamma}(3)\rangle \times D_{\alpha}(1) W^{\alpha}(1)W^{\beta}(2)A_{\gamma}(3)+\mbox{perm.}\nonumber\\
&&F.\quad(D_{\gamma}(3)W^{\alpha}(1))(D_{\alpha}(1)W^{\beta}(2))(D_{\beta}(2)W^{\gamma}(3))+\mbox{perm.}\nonumber\\
&&G.\quad\langle \hat{S}_{ab}(1)\hat{S}_{cd}(2)\rangle \times P_{e}(3)F^{ab}(1)F^{cd}(2)A^{e}(3)+\mbox{perm.}\nonumber\\
&&H.\quad\langle
\hat{S}_{ab}(1)\hat{S}_{cd}(2)\hat{S}_{ef}(3)\rangle \times
F^{ab}(1)F^{cd}(2)F^{ef}(3)
\end{eqnarray}
where the $\hat{S}$ contraction is the usual $\gamma$ trace.

The other contributions, like $(\langle
D_{\alpha}\Omega^{\beta}\rangle P)\cdot
AW^{\alpha}A_{\beta},~~(\langle SS\rangle D_{\alpha})\cdot
FFW^{\alpha},~~(PPD_{\alpha})\cdot AAW^{\alpha}$ and
$(PD_{\alpha}D_{\beta})AWW$, all vanish using $k_{i}\cdot
k_{j}=0,\sla k W=0$ in the Wess-Zumino gauge. We give some details of
the calculation in Appendix \ref{superamplitude}.

Notice that F and H combine to give the GP sum
$1-2+3-4\cdots=1/4$. From these combinations we find the
manifestly supersymmetric 3-point tree amplitudes for vectors and
spinors
\begin{eqnarray}
\label{3pt}
A^{tree}_{3}
&=&k_{1}\cdot A(3)A(1)\cdot A(2)+k_{3}\cdot A(2)A(1)\cdot
A(3)+k_{2}\cdot
A(1)A(2)\cdot A(3)\nonumber\\
&&+iA(1)\cdot W(2)\gamma W(3)+iA(2)\cdot W(3)\gamma
W(1)+iA(3)\cdot W(1)\gamma W(2)\nonumber\\
\end{eqnarray}
where $A(i)$ are the vectors and $W(i)$ the spinors.
(Note that we use the usual anticommuting fields for the spinors; numerical evaluation involves fermionic functional differentiation, replacing these fields with the usual commuting wave functions, and may introduce signs if not all terms have the same ordering.)

This result applies to both the superparticle and superstring.  In
the string case there is also a factor of
$1/(z_{1}-z_{2})(z_{2}-z_{3})(z_{1}-z_{3})$ from the Green
functions, but this is canceled as usual with the inverse factor
from the conformal measure obtained from $\langle
c(1)c(2)c(3)\rangle$.

\mysection{IR regularization}

\mysubsection{Zero modes}

The kinematic factor in supersymmetric amplitudes is closely
related to the spinor zero-mode problem, which is the most
important problem in the Lorentz covariant superparticle and
superstring. If we naively integrate over zero-modes of the
infinite pyramid of spinors with no vertex attached, we find
$0\cdot\infty^{2}\cdot0^{3}\cdot\infty^{4}\cdot0^{5}\cdot\infty^{6}\cdots$.
So we need to regularize the zero-mode integration. In Appendix
\ref{ir} we derive the 2D Green function with a 2D regularization
mass, but it turns out that the zero-mode behavior of the 2D Green
function is exactly that of the 1D one. So we will concentrate on
the 1D case here. To do this IR regularization we introduce small
mass terms in the superparticle free action (for 1D ``proper time" coordinate $z$)
\begin{equation}
X^{a}(-\partial^{2}_z+\xi^{2})X_{a},\quad\quad
-i\pi(\partial_z+\epsilon)\theta
\end{equation}

Now we can fix the measure of zero-modes for $X$ and $\theta$
without ambiguity. For $X$, neglecting the Laplacian term, which
vanishes for zero-modes,
\begin{eqnarray*}
\lim_{\xi\rightarrow0}\int d^D X_0\ e^{-T\xi^{2}X^{2}_0/2 -i(\sum
k)\cdot X_0} &=&
\lim_{\xi\rightarrow0}\left(\frac{2\pi}{T\xi^2}\right)^{D/2}e^{-(\sum
k)^2/2T\xi^2}\\
&=&
(2\pi)^{D}\delta^D\left(\sum k\right)\\
&=& \int d^{D}X_{0}\ e^{-i(\sum k)\cdot X_{0}}
\end{eqnarray*}
where $T$ is the range of $z$ (at 1 loop, the period).  Here we used
$\lim_{\xi\rightarrow0}e^{-x^2/2\xi^2}/\sqrt{\xi}=\sqrt{2\pi}\delta(x)$.
Therefore our zero-mode measure for $X$ is
\begin{equation}
\int d
X_{0}=\lim_{\xi\rightarrow0}\left(\frac{2\pi}{T\xi^2}\right)^{D/2}
\end{equation}
However, this bosonic zero-mode does not appear explicitly, since
this always gives momentum conservation thanks to the vertex
operators.

Similarly for $\theta$ we see
\begin{equation}
\int d\theta\ d\pi~e^{iT\epsilon\pi\theta}=(i\epsilon T)^{\pm
2^{(D-2)/2}}
\end{equation}
where ``$\pm$" stands for fermionic and bosonic spinor
respectively. Then our zero-mode measure for a spinor is
\begin{equation}
\int d\theta\ d\pi=\lim_{\epsilon\rightarrow0}(i\epsilon T)^{\pm
2^{(D-2)/2}}
\end{equation}
In our case we have an infinite pyramid of spinors and hence we
get
\begin{eqnarray}
\int d\theta\ d\pi &=&\lim_{\epsilon\rightarrow0}(i\epsilon T)^{(2^{(D-2)/2})(1-2+3-4+\cdots)}\nonumber\\
&=&\lim_{\epsilon\rightarrow0}(i\epsilon T)^{2^{(D-6)/2}}
\end{eqnarray}
where we used coherent-state regularization for the ambiguous sum
$1-2+\cdots=1/4$:
\begin{eqnarray}
tr[(N+1)(-1)^{N}]&=&\int \frac{d^{2}z}{\pi}e^{-z^{*}z}\langle
z|(a^{\dag}a+1)(-1)^{a^{\dag}a}|z\rangle\nonumber\\
&=& \int
\frac{d^{2}z}{\pi}e^{-|z|^{2}}(\langle z|a^{\dag}a|-z\rangle+\langle z|-z\rangle)\nonumber\\
&=&\int
\frac{d^{2}z}{\pi}(-|z|^{2}+1)e^{-2|z|^{2}}\nonumber\\
&=&-\f{1}{4}+\f{1}{2}=\f{1}{4}
\end{eqnarray}
More intuitively
\begin{eqnarray*}
\frac{1}{1+x}&=&1-x+x^{2}-x^{3}+x^{4}\cdots\\
\frac{1}{(1+x)^{2}}&=&1-2x+3x^{3}-4x^{4}+\cdots
\end{eqnarray*}
so at $x=1$ we get 1/2 and 1/4 respectively.

Therefore in $D=10$ we get effectively $\epsilon^{4}$ for
zero-modes. So our complete spinor measure with non-zero modes is
\begin{equation}
\mathcal{D}\theta\ \mathcal{D}\pi\ (iT\epsilon)^{4}
\end{equation}
The significant role of this effective power will be clear after
we discuss the Green function.

The regularization $1-2+3-4+\cdots=1/4$ explains how we can get a
physical $SO(8)$ spinor contribution out of 2 covariant
16-component spinors $\pi$ and $\theta$. Because we cannot project
covariant spinors into physical spinors in a covariant way, we
need to add infinitely many ghosts to achieve this 1/4 reduction
in amplitudes.

\mysubsection{Regularized Green functions}

We summarize the results of Appendix \ref{ir} here. We find the
regularized 1D Green functions for $x$ and $\theta$
\begin{eqnarray}
\label{worldlinegf} G^{x}(z)&=&\frac{1}{2\xi}\frac{\cosh[\xi
(|z|-T/2)]}{\sinh(\xi T/2)}\nonumber\\
G^{\theta}(z)&=&i(-\partial_{z}+\epsilon)\left[\frac{1}{2\epsilon}\frac{\cosh[\epsilon
(|z|-T/2)]}{\sinh(\epsilon T/2)}\right]
\end{eqnarray}
The $\epsilon$ correction to $\partial_{z}$ in $G^\theta$ is
nontrivial because it multiplies a Green function with a $1/\epsilon$
term.

It is convenient to expand the Green functions in $\epsilon$ when
we calculate scattering amplitudes:
\begin{eqnarray}
G^{x}&=&\frac{1}{\xi^{2}T}+\sum_{n=0}^{\infty}G^{x}_{n}\xi^{n}\nonumber\\
G^{\theta}&=&\frac{i}{\epsilon
T}+\sum_{n=0}^{\infty}G^{\theta}_{n}\epsilon^{n}
\end{eqnarray}
where
\begin{eqnarray}
G^{x}_{0}&=&{T\over 12}+{|z|(|z|-T)\over 2T}
= {T\over 12} +G^x_{un} \nonumber\\
G^\theta_0&=&\frac{i}{2}Sign(z)-i\frac{z}{T} = G^\theta_{un}
\end{eqnarray}
and $G^x_{un}$ and $G^\theta_{un}$ are the usual $1D$
Green functions with periodic boundary conditions, normalized to $G_{un}(0)=0$.
The extra constant will not contribute to massless amplitudes because of derivatives and $k^2=0$.

Because the mass (re)moves zero-modes, the usual fudges of the massless Green functions are eliminated:  There is no freedom to add constants (dependent on $T$, but not $z$) to $G$, and the $\delta$ function in its equation of motion is not modified to $\delta(z)-1/T$ to preserve ``charge conservation".  But the latter property is restored upon expansion in the regulator:
\begin{eqnarray}
G^{x}&=&\frac{1}{\xi^{2}T}+\Delta G^x , \quad
(-\partial^2 +\xi^2)G^x = \delta \quad\Rightarrow\quad
(-\partial^2 +\xi^2)\Delta G^x = \delta -{1\over T}
\nonumber\\
G^{\theta}&=&\frac{i}{\epsilon T}+\Delta G^\theta , \quad
-i(\partial +\epsilon)G^\theta = \delta \quad\Rightarrow\quad
-i(\partial +\epsilon)\Delta G^\theta = \delta  -{1\over T}
\end{eqnarray}

Similarly we will do this expansion for superstring Green
functions. The details are given in Appendix \ref{ir}. However, expansion of
$G^{x}$ is unnecessary, because in vertex operators $X$ appears only as
$\dot{X}$ (and as an argument of the superfields), and any contraction involving this vertex operator is
always finite.  (The derivative kills the potentially divergent
$1/\xi^{2}$ term.)  For this reason $X$ regularization gives only
energy-momentum conservation and is irrelevant to amplitude
corrections. But $\epsilon$ expansion of $G^{\theta}$ is crucial, as
we will see in the next section.

\mysection{Loops}

\mysubsection{\texorpdfstring{N$<$4 super}{N<4 super}}

Here we give simple examples.  The only differences from standard
first-quantization of a loop of a scalar particle or bosonic
string can be associated with ``kinematic factors" that may also
depend on the positions of the vertices on the worldline/sheet.
(For a summary of the standard analysis of the other factors, see
Appendix \ref{loop}.)

Collecting the results of the zero-mode measure and the Green
function zero-mode behavior,
{\it the amplitude is zeroth order in $\epsilon$.}

Since there is an $\epsilon^{4}$ in the measure, we should pick up
an $\epsilon^{-4}$ in the integrand of the path integral. For
example, one sub-diagram of the N-point 1-loop amplitude is
proportional to
\begin{equation}
\oint d\epsilon ~\epsilon^{3}~G^{\theta}(1,2)G^{\theta}(2,3)\cdots
G^{\theta}(N,1)
\end{equation}
Then to evaluate this amplitude we should expand each $G^{\theta}$
and collect terms with $\epsilon^{-4}$.

We now notice that every $G^\theta$
gives $i/\epsilon T$. For $N<4$ there are not enough powers of
$\epsilon^{-1}$ and so their amplitudes just vanish.

\begin{figure}[h]
 \centering
\includegraphics[scale=.5]{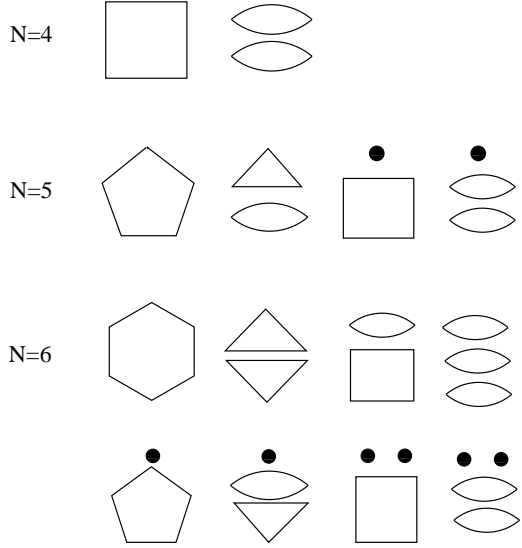}
 \caption{Schematic diagrams for various contractions}
 \label{diagram}
\end{figure}

There is no zero-mode behavior for any contraction involving
$\partial X$ because of the derivative. Therefore $P$ contractions
start to contribute only at $N=5$ (a black dot in Fig.\nobreak\
\ref{diagram}).

\mysubsection{N=4 vector only}

The first nonvanishing amplitude is at $N=4$. However, this is
just the case where every $G^\theta$ from the
$S_{ab}$'s contributes $i/\epsilon T$. So the integration is
trivially done for $K_4$ and only its spin algebra matters.
There are two kinds of diagrams:  the case where all 4 points
are connected, and the case where each pair of points is connected
separately (Fig.\ \ref{diagram}). These two diagrams have opposite
sign. Each closed contraction should be traced over all ghost
pyramid spinors to give $1-2+3-4+\cdots=1/4$. Therefore we get for
$K_{4}$, omitting external field factors,
\begin{eqnarray}
K_{4}&=&-\frac{1}{4}\left[tr(\gamma^{ab}\gamma^{cd}\gamma^{ef}\gamma^{gh})+\mbox{5
permutations}~\right]\nonumber\\
&&+\frac{1}{16}\left[tr(\gamma^{ab}\gamma^{cd})tr(\gamma^{ef}\gamma^{gh})+\mbox{2
permutations}~\right]
\end{eqnarray}

Using the Mathematica code Tracer.m we evaluate this gamma-matrix
trace to find
\begin{eqnarray}
K_{4}&=&~~\f{1}{2}(\delta^{bc}\delta^{de}\delta^{fg}\delta^{ha}+\delta^{be}\delta^{cf}\delta^{dg}\delta^{ha}+\delta^{ae}\delta^{fg}\delta^{ch}\delta^{bd}~+45~ \mbox{terms}\nonumber\\
&&\quad\mbox{from antisymmetrizing each pair of indices}~~[ab][cd][ef][gh])\nonumber\\
&&-\f{1}{2}\left[(\delta^{ac}\delta^{bd}-\delta^{ad}\delta^{bc})(\delta^{eg}\delta^{fh}-\delta^{eh}\delta^{fg})\right.\nonumber\\
&&\quad\quad+(\delta^{ae}\delta^{bf}-\delta^{af}\delta^{be})(\delta^{cg}\delta^{dh}-\delta^{ch}\delta^{dg})\nonumber\\
&&\quad\quad+\left.(\delta^{ag}\delta^{bh}-\delta^{ah}\delta^{bg})(\delta^{ce}\delta^{df}-\delta^{cf}\delta^{de})\right]
\end{eqnarray}
This is the well-known kinematic factor for both tree and 1-loop.
We can also express this results in terms of $F$ as
\cite{bound}
\begin{eqnarray}
&& F^{ac}(1)F^b{}_c(2) F_a{}^d(3) F_{bd}(4)-\frac18 F^{ab}(1)F_{ab}(2) F^{cd}(3)F_{cd}(4)\nonumber\\
&&-\frac14
F^{ab}(1)F^{cd}(2)[F_{ab}(3)F_{cd}(4)-2F_{ac}(3)F_{bd}(4)]
\end{eqnarray}
which can be interpreted as ``graviton", ``dilaton", and ``axion" as far as Lorentz (and not gauge) structure is concerned.  (In the nonplanar case, it actually corresponds to those poles for color singlets in the 1+2=3+4 channel.)

\mysubsection{N=4 super}

Here we again prefer the superfield formalism as explained in
Section 2. However, the 4-point one-loop case is dramatically
simplified due to the IR regularization. Consider the 4 types of
fully contracted operators again and then notice that they can
have only limited $1/\epsilon$ factors, since each $G^\theta$ gives such
a factor while $G^\theta{}'$ doesn't:

\begin{eqnarray}
P_{a}(z_{1})P_{b}(z_{2})&:&\mathcal{O}(\epsilon^{0})\nonumber\\
D_{\alpha}(z_{1})\Omega^{\beta}(z_{2})&:&\mathcal{O}(\epsilon^{0})\nonumber\\
D_{\alpha}(z_{1})D_{\beta}(z_{2})P_{a}(z_{3})&:&\mathcal{O}(\epsilon^{-1})\nonumber\\
D_{\alpha}(z_{1})D_{\beta}(z_{2})D_{\gamma}(z_{3})D_{\delta}(z_{4})&:&\mathcal{O}(\epsilon^{-2})\nonumber\\
(\hat{S})^{n}&:&\mathcal{O}(\epsilon^{-n})
\end{eqnarray}

This means that except for $\hat{S}$ they appear at best from the
6-point at 1 loop.  So the only contractions for this amplitude
are from $\hat S^4$ and $\hat S^{2}d^{2}$.  We also need to
consider the case where 4 $d$'s act on the superfields. Then we
can directly write down the kinematic factor for the manifestly
supersymmetric, 4-point, 1-loop amplitude
\begin{eqnarray}
\label{4pt1loopsuper}
&&\f1{4!}d_{[\alpha}d_{\beta}d_{\gamma}d_{\delta]}
W^{\alpha}(1)W^{\beta}(2)W^{\gamma}(3)W^{\delta}(4)\nonumber\\
&&+\f{3}{32}tr(\gamma^{ab}\gamma^{cd})d_{[\alpha}d_{\beta]}F_{ab}(1)F_{cd}(2)W^{\alpha}(3)W^{\beta}(4)+\mbox{perm.}\nonumber\\
&&+\hat{K}_{4}(F^{4})
\end{eqnarray}
$\hat{K}_{4}$ is the same as
$K_4$ above except that the (super)traces don't include the
physical $\pi,\theta$. Of course, this missing contribution comes
from $(d_{\beta}W^{\alpha})$ $(d_{\alpha}W^{\beta})$
$(d_{\delta}W^{\gamma})$ $(d_{\gamma}W^{\delta})$ plus different
permutations of the $d$'s.  Also, the missing contribution for the
$tr(\gamma^{ab}\gamma^{cd})$ terms comes from
$-\f{1}{4}W^{\alpha}(d_{[\alpha}d_{\delta]}W^{\beta})W^{\gamma}(d_{[\gamma}d_{\beta]})W^{\delta}$
plus different permutations.  Note that this result already has the same form as the 4D N=1 supergraph calculation for N=4 super Yang-Mills \cite{4d} (if we rewrite it in Majorana notation for comparison), where there $tr(I)=4$ already, so $\hat S$ terms are unnecessary to produce $str(I)=16\times 1/4=4$.  (There the $d^4$ comes from overall $\theta$ integration, the $d$'s of the $W$'s being killed by loop-$\theta$ integration.)

We give here the fermion part of the
result of (\ref{4pt1loopsuper}) and leave details to Appendix
\ref{superamplitude}.
\begin{eqnarray}
\label{superresult}
K_{4}^{FFBB}
&=& -\f{i}{2}W(1)\gamma_{ab}\gamma_{c}\partial_{d}W(2)F^{cd}(3)F^{ab}(4)+ 3 \leftrightarrow 4\nonumber\\
&=&\f{i}{2}W(1)\gamma_{abc}\partial_{d}W(2)F^{ab}(3)F^{cd}(4)+iW(1)\gamma_{a}\partial_{b}W(2)F^{ac}(3)F_{c}{}^{b}(4) + 3 \leftrightarrow 4 \nonumber\\
K_{4}^{FFFF}&=&-4k_{1}\cdot k_{4}~W(1)\gamma W(2)\cdot W(3)\gamma
W(4) + 2 \leftrightarrow 4
\end{eqnarray}
where $\gamma_{abc}=\f{1}{3!}\gamma_{[a}\gamma_{b}\gamma_{c]}$.
(The $[abc]$ means to sum over permutations with signs to
antisymmetrize.)
The second form of the FFBB amplitude can be interpreted as ``axion" and ``traceless graviton" terms.  (Using the fermion field equation and symmetry, the former term is totally antisymmetric in $abcd$ and a total curl on the fermions, as the $FF$ factor is then for the bosons, while the latter term is symmetric and traceless in $ab$.)
We have written these amplitudes in manifestly
gauge invariant form.  Note that the complete 4-point amplitude is
totally symmetric in all 4 external lines.  (This was clear from
the original form (\ref{4pt1loopsuper}).)  This means that not
only are the specific cases listed above separately symmetric
between boson lines and between fermion lines (if we had used wave
functions instead of fermionic fields then they would be
antisymmetric), but the amplitudes for other arrangements of
fermions and bosons are obtained simply by permutation. The usual
representations are given in Appendix \ref{superamplitude}.

\mysubsection{\texorpdfstring{N$>$4 vector only}{N>4 vector only}}

In principle there is no difficulty to evaluate higher-point
diagrams. Some new terms occur compared to the $N=4$ case. First
of all, $\partial X$ can contribute from one vertex, acting on a field, which is indicated by a black dot in Fig.\nobreak\ \ref{diagram}.  (All the other vertices contribute contractions between $\theta(z_{i})$ and $\pi(z_{j})$ from $S$.)  Terms of the Green function higher-order in the
$\epsilon$ expansion start to appear and
thus $K_{N}$ has $z_i$ dependence. We give a schematic diagram for
various types of contractions in Fig.\nobreak\ \ref{diagram}.
Notice that our diagram exactly coincides with earlier covariant
RNS results \cite{tsuchiya}. There
can also be corrections from the fermion partition function because of
regularization. For example, this correction in the 6-point amplitude
is proportional to $\theta_{1}'''(0|i\tau)/\theta'_{1}(0|i\tau)$ (see
Appendix \ref{partition}).

\mysubsection{N=5 vector only}

First we will consider the part of the amplitude that doesn't have
a black dot in Fig.\nobreak\ \ref{diagram}. Let's call the graphs
without and with a black dot $K^{a}_{N}$ and $K^{b}_{N}$
respectively. Since the 5-point amplitude has 5 sides we should
choose $G^{\theta}_{0}$ from exactly one side. This is true for
both the pentagon and triangle + ellipse graphs. The difference
between them is the gamma matrix trace factor. So we can write
down the part of $K^{a}_{5}$ for a given group-index ordering (the
$k$th vector has $\theta\gamma^{a_{k}b_{k}}\pi$) as:
\begin{eqnarray}
K^{a}_{5}&=&-\frac{1}{4}[~G^{\theta}_{0}(z_{2}-z_{1})
tr(\gamma^{a_{1}b_{1}}\gamma^{a_{2}b_{2}}\gamma^{a_{3}b_{3}}\gamma^{a_{4}b_{4}}\gamma^{a_{5}b_{5}})+23~\mbox{permutations}~]\nonumber\\
&&+\frac{1}{16}[~G^{\theta}_{0}(z_{2}-z_{1})
tr(\gamma^{a_{1}b_{1}}\gamma^{a_{2}b_{2}}\gamma^{a_{3}b_{3}})tr(\gamma^{a_{4}b_{4}}\gamma^{a_{5}b_{5}})+11~\mbox{permutations}~]\nonumber\\
\end{eqnarray}

Then we can write
\begin{eqnarray}
K^{b}_{5}&=&\sum_{j=2}^{5}k^{a_{1}}_{j}G^x{}'(z_1-z_j)K_{4}(2,3,4,5)+4~\mbox{permutations}\nonumber\\
\end{eqnarray}
where $K_4$ was given in subsection 5.2.  $K^{a}_{5}$ and $K^{b}_{5}$ complete the 5-point planar amplitude.
Totally antisymmetric $\epsilon$-tensor terms vanish because the 5
external momenta are not independent. Notice that the light-cone
GS calculation reduces to our results after heavy algebra
\cite{gs5point}, and RNS needs a spin-structure sum to produce
this result \cite{tsuchiya}.

We postpone the $N\geq6$-point amplitudes to another paper, which
will be interesting because of the anomaly cancellation issue. One
good thing in our covariant formalism is that we have a totally
antisymmetric $\epsilon$-tensor naturally in the hexagon
amplitude, where we have enough momenta to have a nonvanishing
result, contrary to the 5-point case.

\mysection{Future}

There are many avenues of further study, in particular:

\noindent (1)  Many types of diagrams can be calculated.  At the
tree level, diagrams with many fermions have not yet been
explicitly evaluated in any formalism.  New algebraic methods for
the current algebra might be useful.  At the 1-loop level, little
has been done with fermions or higher-point functions.
Alternative IR regularization schemes could be considered.  The
2-loop 4-vector calculation would be a good test, and nothing more
than that has been done at 2 loops, and nothing at all at higher
loops.

\noindent (2)  The Hilbert space needs to be studied covariantly,
especially the vacuum, to completely justify the naive
manipulations we have made for tree graphs.  It would be useful to find the relation of these methods to supergraphs, where explicit zero-mode integrations appear (both in loops, corresponding to $\pi$ zero-modes, and an overall integral for $\theta$ zero-modes.)  Massive vertex
operators for physical states are expected to also be relatively
simple, as the spinor ghosts should appear again in a minimal way
(as opposed to the more complicated structure of the BRST
operator).  The analogy to second-quantized ghost pyramids (e.g., for higher-rank forms) might be useful:  There ghosts beyond the first generation (i.e., the usual Faddeev-Popov ghosts) appear only at 1 loop, to define the measure.

\noindent (3)  Closer relations to other formulations might exist.
An analog to the ${\cal F}_2$ picture of RNS might further
simplify tree calculations.  The many similarities with PS
suggests it might be a particular gauge choice of GP that
truncates the ghost spectrum.

\section*{Acknowledgments}

W.S. thanks Brenno Carlini Vallilo and Nathan Berkovits for
discussions, and Nathan Berkovits for explaining the modern
covariant description of the ${\cal F}_1$ picture.

\appendix

\mysection{Hamiltonian to Lagrangian} \label{hl}

\mysubsection{Superparticle}

In our previous paper we constructed the BRST operator for the
superparticle and superstring in a super Yang-Mills background
\cite{us}. From the BRST operator we can get the gauge fixed
Hamiltonian:
\begin{eqnarray}
H^{particle}_{GF}&=&\{b,Q_{\theta}\}\nonumber\\
&=&-\f{1}{2}\Box+W^{\alpha}\nabla_{\alpha}+\f{1}{2}F^{ab}\theta\gamma_{ba}\pi|_{>}
\end{eqnarray}
where
\begin{eqnarray*}
\Box&=&-(p_{a}+A_{a})^{2},\quad\quad\eta_{ab}=\delta_{ab}\\
\nabla_{\alpha}&=&d_{0\alpha}+A_{\alpha}\\
d_{0\alpha}&=&\pi_{0\alpha}+(\sla p\theta_{0})_\alpha,~~~~
\pi_{0\alpha}=\partial/\partial\theta_{0}^\alpha\\
\gamma^{ab}&=&-\f{1}{4}(\gamma^{a}\gamma^{b}-\gamma^{b}\gamma^{a}),\quad\quad\{\gamma^{a},\gamma^{b}\}=2\delta^{ab}
\end{eqnarray*}
and $\nabla_\alpha, \nabla_a$ are the graded covariant derivatives.

\begin{figure}[h]
 \centering
\includegraphics[scale=.5]{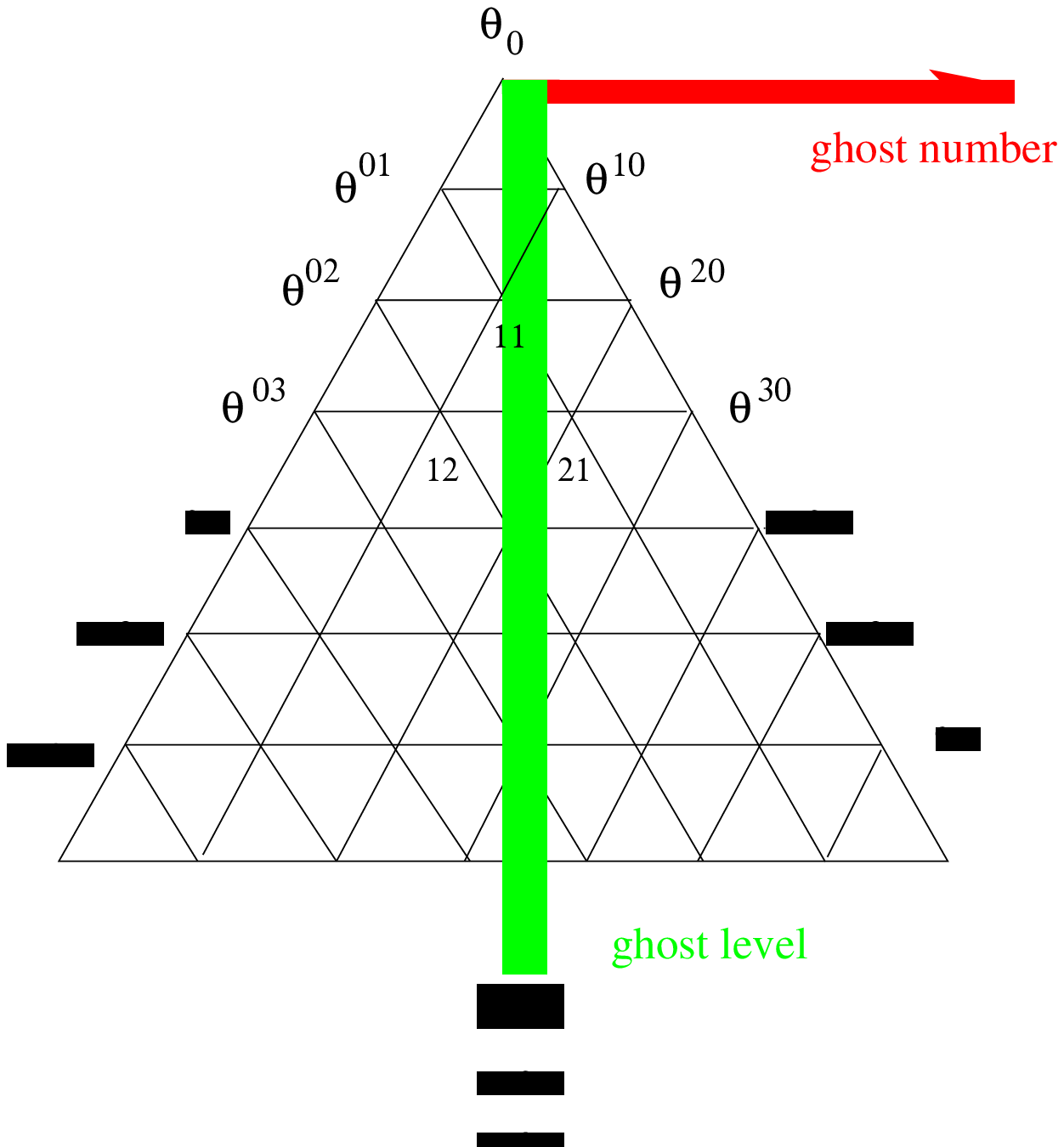}
 \caption{infinite pyramid of ghosts}
\end{figure}

Notice that $\pi$ and $\theta$ are shorthand notation for
$\pi_{p,q}$ and $\theta^{p,q}$, where $p-q$ is the ghost number
and $p+q$ is the ghost level.  (Even level and odd level
correspond to fermion and boson respectively.)  The expression
``$|_{>}$" means ``ghosts only".

Now we go to the Lagrangian form of the action for $x$.
To obtain complete results for the amplitude rules, we need to keep terms in the Hamiltonian quadratic in the background fields.  This has two unusual consequences:  In the Lagrangian, (1) all these terms will become linear (as familiar from the bosonic case), and (2) such terms new to the supersymmetric case will appear only with $\dot\theta$.

Neglecting $ib\dot{c}$ and $F\cdot\hat{s}$ we see
\begin{eqnarray*}
&&-p\cdot
\dot{x}+i\pi\dot{\theta}+\f{1}{2}(p+A)^{2}+W^{\alpha}[\pi_{0\alpha}+p^{a}(\gamma_{a}\theta_{0})_{\alpha}+A_{\alpha}]\\
&\Rightarrow&-\f{1}{2}\dot{x}^{2}+i\pi\dot{\theta}+A\cdot
\dot{x}-iA_{\alpha}\dot{\theta}^{\alpha}+W^{\alpha}[\pi_{0\alpha}+(\dot{x}-A)\cdot(\gamma\theta_{0})_{\alpha}-\f{1}{2}(\gamma\theta)_{\alpha}\cdot
W\gamma\theta]
\end{eqnarray*}
By redefining
$\pi_{0\alpha}\Rightarrow\pi_{0\alpha}+A\cdot(\gamma\theta_{0})_{\alpha}+\f{1}{2}(\gamma\theta_{0})_{\alpha}\cdot
W\gamma\theta_{0}$ (deformed only with gauge fields) we get
\begin{equation*}
-\f{1}{2}\dot{x}^{2}+i\pi\dot{\theta}-A_{\alpha}i\dot{\theta}_{0}^{\alpha}+A\cdot
(\dot{x}+i\theta_{0}\gamma\dot{\theta_{0}})+W^{\alpha}[\pi_{0\alpha}+\dot{x}\cdot(\gamma\theta_{0})_{\alpha}+\f{i}{2}(\gamma\theta_{0})_{\alpha}\cdot
\theta_{0}\gamma\dot{\theta}_{0}]
\end{equation*}
The background terms then give the vertex operator
\begin{eqnarray}
V&=&A^{A}j_{A}\nonumber\\
&=&A_{\alpha}\omega^{\alpha}+A^{a}p_{a}+W^{\alpha}d_{\alpha}+\f{1}{2}F^{ab}\hat{s}_{ba}
\end{eqnarray}
where
\begin{eqnarray}
d_{\alpha}&=&\pi_{0\alpha}+\dot{x}\cdot(\gamma\theta_{0})_{\alpha}+\f{i}{2}(\gamma\theta_{0})_{\alpha}\cdot
\theta_{0}\gamma\dot{\theta}_{0}\nonumber\\
p_{a}&=&\dot{x}+i\theta_{0}\gamma\dot{\theta_{0}}\nonumber\\
\omega^{\alpha}&=&-i\dot{\theta}_{0}^{\alpha}\nonumber\\
\hat{s}_{ab}&=&\theta\gamma_{ab}\pi |_{>}
\end{eqnarray}

The fact that
$\dot{\theta}$ vanishes by its free field equations is related to the fact that its contraction with $\pi$ gives a $\delta(z)$, canceling a (spacetime) propagator, and thus contracting two 3-point vertices into a 4-point vertex.  Thus, they originate from terms in the Hamiltonian quadratic in background fields.  The string vertex operator is the same, with the $z$ derivative replaced with the left- or right-handed worldsheet derivative.

In our previous paper \cite{us} the background coupling had
additional terms involving the expression $R^{a}$, quadratic in
ghost $\theta$'s.  These terms never contribute to amplitudes
because there are no ghost $\pi$'s to cancel them. ($\hat s$ has a
ghost $\pi$, but together with a ghost $\theta$.)  This is also
true for the superstring.

\mysubsection{Superstring}

Like the case of the superparticle, the gauge fixed action for the
superstring comes from $\{\int b,Q_{sstring}\}$, adding
first-order terms: without background,
\begin{eqnarray*}
S_{GF}&=&\int d^{2}z~
\hat{P}^{m}\partial_{m}X-\f{1}{2}\eta_{mn}\hat{P}^{m}\hat{P}^{n}
+i\sqrt{2}\sum_{\pm}\partial_{\pm}c^{\pm}b_{\pm\pm}
+i\sqrt{2}\sum_{\pm}\partial_{\pm}\theta^{\pm}\pi^{\pm}
\end{eqnarray*}
The $\sqrt{2}$ comes from
$\partial_{\pm}=(1/\sqrt{2})(\partial_{0}\pm\partial_{1})$.

We can introduce the background as for the particle case:
\begin{eqnarray}
V&=&A^{A}J_{A}\nonumber\\
&=&A_{\alpha}\Omega^{\alpha}+A^{a}P_{a}+W^{\alpha}D_{\alpha}+\f{1}{2}F^{ab}\hat{S}_{ba}
\end{eqnarray}
where
\begin{eqnarray}
D_{\alpha}&=&\pi_{0\alpha}+(\gamma^{a}\theta_{0})_{\alpha}\partial X_{a}+ i\f{1}{2}(\gamma^{a}\theta_{0})_{\alpha}\theta_{0}\gamma_{a}\partial\theta_{0}\nonumber\\
P_{a}&=&\partial X_{a}+ i\theta_{0}\gamma_{a}\partial\theta_{0}\nonumber\\
\Omega_{\alpha}&=&-i\partial\theta_{0\alpha}\nonumber\\
\hat{S}_{ab}&=&\theta\gamma_{ab}\pi |_{>}
\end{eqnarray}

\mysection{Current algebra} \label{ca}

The operator (affine Lie) algebra remains simple because the
currents are no more than cubic in the fundamental variables:
\begin{equation}
J_{A}(z_{1})J_{B}(z_{2})=G_{AB}'(z_{1}-z_{2}){f_{AB}}^{C}[z']J_{C}(z')+G_{AB}''(z_{1}-z_{2})\mbox{\boldmath$\eta$}_{AB}+::J_{A}(z_{1})J_{B}(z_{2})::
\end{equation}
where $J_A$ has zero-modes $j_A$, of which only $p_a$ and
$d_\alpha$ act nontrivially on $A^A$, and $G_{AB}$ is the relevant
Green function. For example, $G^{(n)}_{ab}=G^{(n)}_{x}$ and
$G^{(n)}_{\alpha\beta}=G^{(n-1)}_{\theta}$. The various
definitions are

\begin{eqnarray*}
\label{calgebra}
{f_{\alpha\beta}}^{a}[z]P_{a}(z)&\equiv&\gamma^{a}_{\alpha\beta}[P_{a}(z_{\alpha})+P_{a}(z_{\beta})]\\
f_{ a\alpha\beta}[z]\Omega^\beta(z)&\equiv&2\gamma_{a\alpha\beta}\Omega^{\beta}(z_{a})\\
f_{\alpha a\beta}[z]\Omega^\beta(z)&\equiv&-2\gamma_{a\alpha\beta}\Omega^{\beta}(z_{a})\\
\mbox{\boldmath$\eta$}_{\alpha}{}^{\beta}&=&-i\delta_{\alpha}^{\beta}\\
\mbox{\boldmath$\eta$}^{\beta}{}_{\alpha}&=&i\delta_{\alpha}^{\beta}\\
\mbox{\boldmath$\eta$}_{a b}&=&-\eta_{a b}\\
&&\mbox{otherwise vanish}
\end{eqnarray*}
and
\begin{eqnarray}
::J_{A}(z_{1})J_{B}(z_{2})::A^{C}&\equiv&:J_{A}(z_{1})J_{B}(z_{2}):A^{C}\nonumber\\
J_A(z_{1}) A^B(z_{2})&=& G_{AB}'(z_{1}-z_{2})(j_A A^B)(z_{2})
+\normalord{ J_A(z_{1})
A^B(z_{2})}\nonumber\\
A^A(z_{1}) A^B(z_{2})&=&e^{-k_1\cdot
k_2G_{x}(z_{1}-z_{2})}\normalord{A^A(z_{1}) A^B(z_{2})}\nonumber\\
::J_{A}(z_{1})J_{B}(z_{2})::J_{C}(z_{3})&\equiv&(-1)^{BC}G_{AC}'(z_{1}-z_{3}){f_{AC}}^{D}[z']J_{D}(z')J_{C}(z_{3})\nonumber\\
&&+G_{BC}'(z_{2}-z_{3}){f_{BC}}^{D}[z']J_{A}(z_{1})J_{D}(z')\nonumber\\
&&+(-1)^{BC}G_{AC}'(z_{1}-z_{3}){f_{AC}}^{D}[z']::J_{D}(z')J_{C}(z_{3})::\nonumber\\
&&+G_{BC}'(z_{2}-z_{3}){f_{BC}}^{D}[z']::J_{A}(z_{1})J_{D}(z')::\nonumber\\
&&+::J_{A}(z_{1})J_{B}(z_{2})J_{C}(z_{3})::
\end{eqnarray}
It is then straightforward to get (\ref{simplerule}), which is all
that is needed in amplitude calculations.

\mysection{Component expansions} \label{ce}

The $\theta$ expansion of the superfields follows directly from
the constraints on the (super)field strengths
\begin{eqnarray*}
\label{algebra}
~[\nabla_{a},\nabla_{b}]~&=&~F_{ab} \\
\{\nabla_{0 \alpha},\nabla_{0 \beta}\}~&=&~2\gamma_{a \alpha
\beta} \nabla^{a}\\
~[\nabla_{0 \alpha},\nabla_{a}]~&=&2\gamma_{a \alpha
\beta}W^{\beta}
\end{eqnarray*}
and the Bianchi identities that follow from them.

Although in practice we perform component expansions by evaluating
spinor derivatives at $\theta=0$, we can also directly expand
superfields in $\theta$.  In a Wess-Zumino gauge we have:
\begin{eqnarray*}
F_{ab}~&=&~\on\circ F{}_{ab}\\
W^{\alpha}&=&\on\circ W{}^{\alpha}+~\f{1}{2}(\gamma^{ab}\theta_{0})^{\alpha}\on\circ F{}_{ab}\\
A_{a}&=&\on\circ A{}_a~+~2\theta_{0}\gamma_{a}\on\circ
W+~\f{1}{2}\theta_{0}\gamma_{a}\gamma^{bc}\theta_{0}~\on\circ
F{}_{bc}\\
A_{\alpha}&=&~(\gamma^{a}\theta_{0})_{\alpha}\on\circ
A{}_a~+~\f{4}{3}(\gamma^{a}\theta_{0})_{\alpha}\theta_{0}\gamma_{a}\on\circ
W+~\f{1}{4}(\gamma^{a}\theta_{0})_{\alpha}\theta_{0}\gamma_{a}\gamma^{bc}\theta_{0}\on\circ
F{}_{bc}\qquad
\end{eqnarray*}
where $\circ$ indicates $\theta_{0}$ independence, and we have
expanded only to constant field strengths $W$ and $F$, which is
sufficient for lower-point diagrams because of the deficiency of
$\pi$'s. The vertex operator $V=V_B+V_F$ for the superparticle is
then :
\begin{eqnarray}
V_B&=& i\on\circ A\cdot\dot{x}+\f{1}{2}\on\circ F{}^{ab}\theta\gamma_{ba}\pi\nonumber\\
V_F&=&\on\circ W
{}^{\alpha}[\pi_{0\alpha}-i\dot{x}\cdot(\gamma\theta_{0})_{\alpha}-\f{i}{6}(\gamma_{a}\theta_{0})_{\alpha}\theta_{0}\gamma^{a}\dot{\theta}_{0}]
\end{eqnarray}
Here $\theta\gamma_{ab}\pi $ includes the physical
$\pi_{0},\theta_{0}$.  (In terms of superfields and currents we
hide this physical $\pi_{0},\theta_{0}$ in $W^{\alpha}$ and
$D_{\alpha}$. Then $\hat{S}$ has only ghost number non-zero
$\pi,\theta$.) Notice that the spinor vertex is the supersymmetry
generator $q_{\alpha}$, which will happen again in the superstring
case. Inserting plane waves for the fields,
\begin{eqnarray}
V_{B}&=&A_{a}(i\dot{x}^{a}+\theta\gamma^{ab}\pi k_{b})e^{ik\cdot x}\\
V_{F}&=&
w^{\alpha}[\pi_{0\alpha}-i\dot{x}\cdot(\gamma\theta_{0})_{\alpha}-\f{i}{6}(\gamma_{a}\theta_{0})_{\alpha}\theta_{0}\gamma^{a}\dot{\theta}_{0}]e^{ik\cdot
x}
\end{eqnarray}
The superstring vertices are essentially the same.

\mysection{Loop review} \label{loop}

\mysubsection{Superparticle}

Since our theory is 1st-quantized, we should calculate amplitudes
in terms of worldline Green functions with periodic boundary
conditions \cite{strassler}. So our partition function with
imaginary time is
\begin{equation}
\mathcal{N}\int_{0}^{\infty}\frac{dT}{T}\int\mathcal{D}X~\mathcal{D}c~\mathcal{D}b~\mathcal{D}\theta~\mathcal{D}\pi
~Tr ~e^{-\int^{T}_{0}dz\ L_{\theta}}
\end{equation}
where
\begin{equation}
\label{normailzation} \mathcal{N}=\int\mathcal{D}P\
e^{-\int^{T}_{0} dz P^{2}/2}
\end{equation}
and $\int dT/T$ comes from the Schwinger proper-time integral
representation of the 1-loop vacuum energy $-Tr[\ln(-\Box)]$.

Since this is a 1-loop amplitude, we should impose periodic
boundary conditions, on both $X$ and $\theta$ (to preserve
supersymmetry):
\begin{equation}
X(T)=X(0),\quad\quad\theta(T)=\theta(0)
\end{equation}
This boundary condition also results in a supertrace naturally in
the loop amplitude.

In this setting the color-ordered, N-point, 1-loop amplitude of
the superparticle can be written as:
\begin{equation}
\label{particleamplitude} A_{N}= G_{N}\int_{0}^{\infty}dT\ (2\pi
T)^{-D/2} \int_{0\le z_r\le z_{r+1}\le z_N=T\le\infty}d^{N-1}z_i\
K_{N}\ e^{-\sum_{1\leq r<s\leq N} k_r\cdot k_s G_{rs}}
\end{equation}
The factor $(2\pi T)^{-D/2}$ comes from
$\mathcal{N}\int\mathcal{D}Xe^{-\int\dot{X}^{2}/2}$. The $z_i$
integration factors come from the Nth order expansion of the
vertex operator. The worldline Green function $G(z_s-z_r)$ is
given in (\ref{worldlinegf}).  Examples of the kinematic factor
$K_{N}$ are given in section 5.  The factor $G_{N}$ is the trace of group
generators in a given ordering.

\mysubsection{Superstrings}
\label{partition}

The procedure is almost identical to the particle case. One
difference is that our Green functions are now doubly periodic:
\begin{eqnarray}
&&G^x(z)= G^x(z+2\pi i)= G^x(z+T)\nonumber\\
&&G^\theta(z)= G^\theta(z+2\pi i)= G^\theta(z+T)
\end{eqnarray}
Again this periodic boundary condition in both directions is
required by supersymmetry. Also, there is a topological
distinction among graphs, namely planar, nonplanar, and
unorientable graphs.  We will concentrate on the planar one here;
the others follow from similar considerations.

We can write the color-ordered, N-point, 1-loop, superstring
amplitude in a form identical to that of the particle case
(\ref{particleamplitude}), but with the string Green function
given in Appendix \ref{ir}. After the usual change of variables
\begin{equation}
\rho_i = e^{-z_i},\quad\quad w = \rho_{N}=e^{-T}
\end{equation}
we have
\begin{equation}
G_N\int_{0}^{1} \frac{dw}{w}\left(\frac{-2\pi}{\ln w}\right)^{D/2}
\int_{0\le w\le\rho_{r+1}\le\rho_r\le
1}\prod_{r=1}^{N-1}\frac{d\rho_{r}}{\rho_{r}}\ K_{N}\
e^{-\sum_{1\leq r<s\leq N} k_r\cdot k_s G_{rs}}
\end{equation}

In the bosonic case there is a factor of $[f(w)]^{-D+2}$ coming
from the partition function (and a $w$ from the
tachyon mass) for $D$ $X$'s and the 2 reparametrization ghosts
$b$ and $c$. In the supersymmetric case this is canceled (as in
all superstring formulations) by an
$[f(w)]^{(2^{D/2})(1-2+3-4+\cdots)}=[f(w)]^{2^{(D-4)/2}}$ which
comes from the infinite pyramid of spinors, in $D=10$. However,
the regularization introduces corrections to the spinor partition function:
\begin{equation}
\prod_{n}(1-w^{n})^{4}(1-w^{n})^{4}\Rightarrow\prod_{n}(1-w^{n+i\epsilon})^{4}(1-w^{n-i\epsilon})^{4}
\end{equation}
The $\epsilon$ expansion of this partition function gives
corrections to amplitudes. For example, in the 6-point, 1-loop amplitude
we expect a term
$\sim(1/\epsilon)^{6}(\epsilon^{2}\theta_{1}'''/\theta'_{1})$, where the
$(1/\epsilon)^{6}$ comes from 6 $G^\theta$'s and
the $\epsilon^{2}\theta_{1}'''/\theta'_{1}$ comes from expansion of the
spinor partition function.

\mysection{Periodic Green functions} \label{ir}

\mysubsection{Second order}

The general Fourier decomposition of a function in 2 dimensions
with doubly periodic boundary conditions
$((x,y)\simeq(x+2\pi,y)\simeq(x,y+2\pi\tau))$ for real $\tau=T/2\pi$ is
\begin{equation}
G(x-x',y-y')=\sum_{n,m}G_{n,m}~e^{in(x-x')+im(y-y')/\tau}
\end{equation}
Then the $G_{n,m}$ for the Green function of the differential
operator $-\partial^{2}_{x}-\partial^{2}_{y}+\epsilon^{2}$ is
easily found to be
\begin{equation}
G_{n,m}=\frac{1}{2\pi\tau}\frac{1}{n^{2}+\frac{m^{2}}{\tau^{2}}+\epsilon^{2}}
\end{equation}
For simplicity we can set $x'=y'=0$ by translational invariance.
Using Schwinger proper-time parametrization we get
\begin{equation}
G(x,y)=\frac{1}{2\pi\tau}\sum_{n,m}\int_{0}^{\infty} ds~
e^{-s(n^2+m^2/\tau^2+\epsilon^2)+inx+imy/\tau}
\end{equation}
Next using Jacobi's transform
\begin{equation*}
\sum_{m}~e^{-sm^2/\tau^2+imy/\tau}=\tau\sqrt{\frac{\pi}{s}}\sum_{m}~
e^{-(2\pi m-y/\tau)^2\tau^2/4s}
\end{equation*}
we get
\begin{eqnarray}
G(x,y)=\frac{1}{2\pi}\sum_{n,m}\int_{0}^{\infty}
ds~\sqrt{\frac{\pi}{s}}~e^{-(2\pi
m-y/\tau)^2\tau^2/4s-s(n^{2}+\epsilon^{2})+inx}
\end{eqnarray}
Then using
\begin{equation*}
\int^{\infty}_{0}ds~s^{\alpha-1}~e^{-ps-q/s}=
2\left(\frac{q}{p}\right)^{\alpha/2}K_{\alpha}(2\sqrt{pq})
\end{equation*}
we get
\begin{equation}
G(x,y)=\frac1{\sqrt{2\pi}}\sum_{n,m}\left(\frac{(2\pi
m\tau-y)^{2}}{n^{2}+\epsilon^{2}}\right)^{1/4}e^{inx}K_{1/2}\left(\sqrt{(2\pi
m\tau-y)^{2}(n^{2}+\epsilon^{2})}\right)
\end{equation}
Also using $K_{1/2}(z)=\sqrt{\pi/2z}e^{-z}$ we get
\begin{eqnarray}
G(x,y)&=&\frac12\sum_{n,m}\frac{1}{\sqrt{n^{2}+\epsilon^{2}}}e^{-|2\pi
m\tau-y|\sqrt{n^{2}+\epsilon^{2}}+inx}\nonumber\\
&=&\frac{1}{2\epsilon}e^{-\epsilon
|y|}+\frac{1}{2\epsilon}\sum_{m\neq0}e^{-\epsilon|2\pi
m\tau-y|}+\frac12\sum_{n\neq0}\frac{1}{\sqrt{n^{2}+\epsilon^{2}}}e^{-|y|\sqrt{n^{2}+\epsilon^{2}}+inx}\nonumber\\
&&\frac12\sum_{n,m\neq0}\frac{1}{\sqrt{n^{2}+\epsilon^{2}}}e^{-|2\pi
m\tau-y|\sqrt{n^{2}+\epsilon^{2}}+inx}
\end{eqnarray}
Now let's transform each sum into a sum over positive integers
only:
\begin{eqnarray}
\frac{1}{2\epsilon}\sum_{m\neq0}e^{-\epsilon|2\pi
m\tau-y|}&=&\frac1\epsilon\cosh(\epsilon
y)\sum^{\infty}_{m=1}e^{-2\pi m\tau\epsilon}\nonumber\\
&=&\frac1{2\epsilon}e^{-\pi\tau\epsilon}\frac{\cosh(\epsilon
y)}{\sinh(\pi\tau\epsilon)}
\end{eqnarray}
\begin{equation}
\frac12\sum_{n\neq0}\f{1}{\sqrt{n^{2}+\epsilon^{2}}}e^{-|y|\sqrt{n^{2}+\epsilon^{2}}~+~inx}
=\frac12\sum_{n=1}^{\infty}\f{1}{\sqrt{n^{2}+\epsilon^{2}}}(\lambda_{n}^{n}+c.c.)
\end{equation}
\begin{equation}
\frac12\sum_{n,m\neq0}\f{1}{\sqrt{n^{2}+\epsilon^{2}}}e^{-|2\pi
m\tau-y|\sqrt{n^{2}+\epsilon^{2}}~+~inx}=\frac12\sum_{n,m=1}^{\infty}\f{1}{\sqrt{n^{2}+\epsilon^{2}}}
w^{mn}_{n}(\rho_{n}^{n}+\rho_{n}^{-n}+c.c.)
\end{equation}
where
\begin{eqnarray}
\ln\lambda_{n}&=&i\left(x+i|y|\sqrt{1+\frac{\epsilon^{2}}{n^{2}}}\right)\nonumber\\
\ln\rho_{n}&=&i\left(x+iy\sqrt{1+\frac{\epsilon^{2}}{n^{2}}}\right)\nonumber\\
w_{n}&=&e^{-2\pi\tau\sqrt{1+\epsilon^2/n^2}}\nonumber\\
\end{eqnarray}
We can subtract out $G$ for the particle from $G(x,y)$, which includes the part divergent as $\epsilon\rightarrow0$:
\begin{equation}
\frac1{2\epsilon}\frac{\cosh[\epsilon
(|y|-\pi\tau)]}{\sinh(\pi\tau\epsilon)}
\end{equation}
The remainder is
\begin{equation}
\frac12\sum_{n=1}^{\infty}\f{1}{\sqrt{n^{2}+\epsilon^{2}}}(\lambda_{n}^{n}+c.c.)+\frac12\sum_{n,m=1}^{\infty}\f{1}{\sqrt{n^{2}+\epsilon^{2}}}
w^{mn}_{n}(\rho_{n}^{n}+\rho_{n}^{-n}+c.c)
\end{equation}
In the $\epsilon\rightarrow0$ limit
($w_{n}\rightarrow w=e^{-T},
\lambda_{n}\rightarrow\lambda,
\rho_{n}\rightarrow\rho=e^{-z},
z=y-ix$), using
\begin{equation*}
\sum_{n=1}^{\infty}\frac{x^{n}}{n}=-\ln(1-x)\\
\end{equation*}
we get for the remainder
\begin{eqnarray*}
&&-\ln|1-\lambda|-\sum_{m=1}^{\infty}\ln|(1-w^{m}\rho)(1-w^{m}\rho^{-1})|\\
&&\quad\quad=-\frac{[Re(z)]^2}{2T}-\frac12\ln|\lambda|-\ln|f(w)^{2}|+G^x_{un}
\end{eqnarray*}
where (assuming $|y|=y$ for simplicity)
\begin{eqnarray*}
G^x_{un}(z,T)&=&-\ln\left|\frac{2\pi\theta_1(\frac{iz}{2\pi
}|\f{iT}{2\pi})}{\theta'_1(0|\f{iT}{2\pi})}\right|+\frac{[Re(z)]^2}{2T}\\
\theta_1(\f{iz}{2\pi}|\f{iT}{2\pi})&=&-iw^{1/8}(\rho^{1/2}-\rho^{-1/2})
\prod_{m=1}^\infty(1-w^m\rho)(1-w^m\rho^{-1})(1-w^{m})\\
\theta'_1(0|\f{iT}{2\pi})&=&2\pi w^{1/8}f^3(w),\qquad
f(w)=\prod_{m=1}^\infty(1-w^m)\\
\end{eqnarray*}
and $G^x_{un}$ is the unregularized Green function with the usual $T$-dependent ``constant" added to normalize its short distance behavior to be the same as that of the tree case (see, e.g., \cite{fields}).

Combining the two parts we get
\begin{eqnarray}
G(\rho)&=&\frac1{T\epsilon^{2}}+\frac{y^{2}}{2T}-\frac{y}2+\frac{T}{12}+\mathcal{O}(\epsilon)\nonumber\\
&&+\frac12 Re\left(\frac{\ln^{2}\rho}{\ln
w}\right)-\frac12\ln|\rho|-\ln|[f(w)]^{2}|+G^x_{un}+\mathcal{O}(\epsilon)\nonumber\\
&=&\frac1{T\epsilon^{2}}-{1\over 12}\ln
|w[f(w)]^{24}|+G^x_{un}+\mathcal{O}(\epsilon)
\end{eqnarray}
The first term is the zero-mode behavior, and the second term is a constant that won't contribute to massless amplitudes (because of derivatives and $k^2=0$; the non-$f$ piece is the same as for the particle).

\mysubsection{First order}

The worldsheet Green function for $\theta$ can be obtained by
differentiating that for $X$. However, to be careful about
zero-modes some modification is needed. For the 1st-order
differential operator $-i(\partial_{y}-i\partial_{x}+\epsilon)$ we
find the mode sum of the Green function
\begin{eqnarray*}
G^\theta&=&{i\over T}\sum_{m,n}\frac{-im/\tau+n+\epsilon}{m^2/\tau^2+(n+\epsilon)^{2}}e^{inx+ imy/\tau}\\
&=&i(-\partial_{y}-i\partial_{x}+\epsilon)\frac1{2\pi\tau}\sum_{m,n}\frac{1}{m^2/\tau^2
+(n+\epsilon)^{2}}e^{inx+imy/\tau}
\end{eqnarray*}
where the $\epsilon$ in the numerator is nontrivial
because the second-order Green function has a
$1/\epsilon$ pole. The above sum is almost identical to the
second-order case except for the change
$n^{2}+\epsilon^{2}\Rightarrow (n+\epsilon)^{2}$. Therefore we can
write the first-order Green function as
\begin{eqnarray}
G^{\theta}&=&i(-\partial_{y}-i\partial_{x}+\epsilon)\left[\frac1{2\epsilon}\frac{\cosh[\epsilon
(|y|-\pi\tau)]}{\sinh(\pi\tau\epsilon)}+\frac12\sum_{n=1}^{\infty}\frac{1}{n+\epsilon}(\lambda_{n}^{n}+c.c.)\right.\nonumber\\
&&\left.\hspace{1cm}+\frac12\sum_{n,m=1}^{\infty}\frac{1}{n+\epsilon}
w^{mn}_{n}(\rho_{n}^{n}+\rho_{n}^{-n}+c.c.)\right]
\end{eqnarray}
where
\begin{eqnarray*}
\ln\lambda_{n}&=&i\left[x+i|y|\left(1+\frac{\epsilon}{n}\right)\right]\\
\ln\rho_{n}&=&i\left[x+iy\left(1+\frac{\epsilon}{n}\right)\right]\\
w_{n}&=&e^{-2\pi\tau(1+\epsilon/n)}
\end{eqnarray*}
Hence it has a less divergent leading term followed by the
expected differentiated second-order Green function:
\begin{eqnarray}
G^\theta&=&{i\over T\epsilon}+\sum_{n=0}^{\infty}G^\theta_n\epsilon^n
\nonumber\\
&=&{i\over T\epsilon}+i(\partial_{y}+i\partial_{x})G^x_0+\mathcal{O}(\epsilon)
\end{eqnarray}

\mysection{Super amplitudes}
\label{superamplitude}

\mysubsection{Super tree}

In this
section we give some details of the calculation of 3-point tree and 4-point 1-loop super amplitudes.

We will concentrate on terms which give
fermion contributions. In (\ref{supertree}) only $C,D,$ and $E$
give the $AWW$ amplitude. Let's consider $A(1)W(2)W(3)$, for example.
Using (\ref{simplerule}),(\ref{calgebra}) we get for the tree (no
fermion zero-mode regularization)
\begin{eqnarray}
\langle P_{a}(1)P_{b}(2)\rangle&=&-\eta_{ab}\f{1}{(z_{1}-z_{2})^{2}}\nonumber\\
\langle D_{\alpha}(1)\Omega^{\beta}(2)\rangle&=&\delta_{\alpha}^{\beta}\f{1}{(z_{1}-z_{2})^{2}}\nonumber\\
\langle P_{a}(1)D_{\alpha}(2)D_{\beta}(3)\rangle&=&-i\gamma_{a\alpha\beta}[\f{2}{z_{1}-z_{2}}\f{1}{(z_{1}-z_{3})^{2}}-\f{2}{z_{1}-z_{3}}\f{1}{(z_{2}-z_{1})^{2}}\nonumber\\
&&-\f{1}{z_{2}-z_{3}}(\f{1}{(z_{1}-z_{2})^{2}}+\f{1}{(z_{1}-z_{3})^{2}})]
\end{eqnarray}
Then we see in the Wess-Zumino gauge
($A_{\alpha}=\gamma^{a}_{\alpha\beta}\theta^{\beta}A^{a}+\mathcal{O}(\theta^{2})+\cdots$, etc.)
\begin{eqnarray}
C:(PPD)&:&-\langle P_{a}(1)P_{b}(2)\rangle
A^{a}(1)(D_{\alpha}(3)A^{b}(2))W^{\alpha}(3)\nonumber\\
&=&-i\f{1}{(z_{1}-z_{2})^{2}}\f{1}{z_{3}-z_{2}}A^{a}(1)(d_{\alpha}A_{a}(2))W^{\alpha}(3)\nonumber\\
&=&-i\f{1}{(z_{1}-z_{2})^{2}}\f{2}{z_{3}-z_{2}}A^{a}(1)W(2)\gamma_{a}W(3)\nonumber\\
(PDP)&:&-\langle P_{a}(1)P_{b}(3)\rangle
A^{a}(1)(D_{\alpha}(2)A^{b}(3))W^{\alpha}(2)\nonumber\\
&=&-i\f{1}{(z_{1}-z_{3})^{2}}\f{2}{z_{2}-z_{3}}A^{a}(1)W(3)\gamma_{a}W(2)\nonumber\\
D:(PDD)&:&-\langle
P_{a}(1)D_{\alpha}(2)D_{\beta}(3)\rangle A^{a}(1)W^{\alpha}(2)W^{\beta}(3)\nonumber\\
&=&i[\f{2}{z_{1}-z_{2}}\f{1}{(z_{1}-z_{3})^{2}}-\f{2}{z_{1}-z_{3}}\f{1}{(z_{2}-z_{1})^{2}}\nonumber\\
&&-\f{1}{z_{2}-z_{3}}(\f{1}{(z_{1}-z_{2})^{2}}+\f{1}{(z_{1}-z_{3})^{2}})]A(1)\cdot
W(2)\gamma W(3)\nonumber\\
E:(\Omega DD)&:&\langle\Omega^{\alpha}(1)D_{\beta}(2)\rangle
(D_{\gamma}(3)A_{\alpha}(1))W^{\beta}(2)W^{\gamma}(3)\nonumber\\
&&-\langle\Omega^{\alpha}(1)D_{\gamma}(3)\rangle
(D_{\beta}(2)A_{\alpha}(1))W^{\beta}(2)W^{\gamma}(3)\nonumber\\
&=&i\left[-\f{1}{(z_{1}-z_{2})^{2}}\f{1}{z_{3}-z_{1}}+\f{1}{(z_{1}-z_{3})^{2}}\f{1}{z_{2}-z_{1}}\right]A(1)\cdot
W(2)\gamma W(3)
\end{eqnarray}
This reduces to (\ref{3pt}).

 \mysubsection{1 loop : 2 fermions + 2 vectors}

We will concentrate on the case where the fermions are
at both ends. The other case can be easily obtained by
permutation. There are two kinds of contributions: $W(dW)(dddW)W$
(with $WF(ddF)W$) and $W(dW)(dW)(ddW)$ (and
corresponding $W(dW)F(dF)$). The $W^{2}F^{2}$ contribution gives a
GP sum with the corresponding $W^2(dW)^2$ as usual. The explicit formula is
\begin{eqnarray}
A:&-&W^{\alpha}(1)\left(d_{\gamma}W^{\beta}(2)\right)\left(\f{1}{3!}d_{[\delta}d_{\alpha}d_{\beta]}W^{\gamma}(3)\right)W^{\delta}(4)\nonumber\\
&+&\f{3}{16}tr(\gamma_{ab}\gamma_{cd})W^{\alpha}(1)F^{ab}(2)\left(\f{1}{2!}d_{[\delta}d_{\alpha]}F^{cd}(3)\right)W^{\delta}(4)\nonumber\\
B:&-&W^{\alpha}(1)\left(\f{1}{3!}d_{[\delta}d_{\alpha}d_{\gamma]}W^{\beta}(2)\right)\left(d_{\beta}W^{\gamma}(3)\right)W^{\delta}(4)\nonumber\\
&+&\f{3}{16}tr(\gamma_{ab}\gamma_{cd})W^{\alpha}(1)\left(\f{1}{2!}d_{[\delta}d_{\alpha]}F^{ab}(2)\right)F^{cd}(3)W^{\delta}(4)\nonumber\\
C:&-&W^{\alpha}(1)\left(d_{\alpha}W^{\beta}(2)\right)\left(d_{\delta}W^{\gamma}(3)\right)\left(\f{1}{2!}d_{[\beta}d_{\gamma]}W^{\delta}(4)\right)\nonumber\\
&+&\f{3}{16}tr(\gamma_{ab}\gamma_{cd})W^{\alpha}\left(d_{\alpha}W^{\beta}(2)\right)F^{ab}(3)\left(d_{\beta}F^{cd}(4)\right)\nonumber\\
D:&-&W^{\alpha}(1)\left(d_{\delta}W^{\beta}(2)\right)\left(d_{\alpha}W^{\gamma}(3)\right)\left(\f{1}{2!}d_{[\gamma}d_{\beta]}W^{\delta}(4)\right)\nonumber\\
&+&\f{3}{16}tr(\gamma_{ab}\gamma_{cd})W^{\alpha}(1)F^{ab}(2)\left(d_{\alpha}W^{\gamma}(3)\right)\left(d_{\gamma}F^{cd}(4)\right)\nonumber\\
E:&&\left(d_{\beta}d_{\gamma}W^{\alpha}(1)\right)\left(d_{\delta}W^{\beta}(2)\right)\left(d_{\alpha}W^{\gamma}(3)\right)W^{\delta}(4)\nonumber\\
&-&\f{3}{16}tr(\gamma_{ab}\gamma_{cd})\left(d_{\beta}F^{ab}(1)\right)\left(d_{\delta}W^{\beta}(2)\right)F^{cd}(3)W^{\delta}(4)\nonumber\\
F:&&\left(d_{\gamma}d_{\beta}W^{\alpha}(1)\right)\left(d_{\alpha}W^{\beta}(2)\right)\left(d_{\delta}W^{\gamma}(3)\right)W^{\delta}(4)\nonumber\\
&-&\f{3}{16}tr(\gamma_{ab}\gamma_{cd})\left(d_{\gamma}F^{ab}(1)\right)F^{cd}(2)\left(d_{\delta}W^{\gamma}(3)\right)W^{\delta}(4)
\end{eqnarray}
$A$ and $B$ vanish due to a GP sum. $C+D$ and $E+F$ give
identical contributions, using integration by parts (momentum conservation) and the (free) $W$ field equation $\sla\partial W=0$.  The results are given in (\ref{superresult}).

These results appear in the literature in forms where neither gauge invariance nor permutation symmetry (relating FFBB and FBFB) is manifest, which we now provide for comparison.
When written in terms of each momentum and gauge field,
the results are (before applying integration by parts)
\begin{eqnarray}
\label{cdef} C:&&-\f{1}{2}k_{1}\cdot k_{2} W(1)\sla A_{2}\sla
k_{2} \sla A_{3}W(4) +\f{1}{2}A_{3}\cdot k_{4} W(1)\sla
A_{2}\sla
k_{2} \sla k_{3}W(4)\nonumber\\
D:&&-\f{1}{2}k_{1}\cdot k_{3} W(1)\sla A_{3}\sla k_{3} \sla
A_{2}W(4) +\f{1}{2}A_{2}\cdot k_{4} W(1)\sla A_{3}\sla
k_{3} \sla k_{2}W(4)\nonumber\\
E:&&-\f{1}{2}k_{1}\cdot k_{3} W(4)\sla A_{2}\sla k_{2} \sla
A_{3}W(1) +\f{1}{2}A_{3}\cdot k_{1} W(4)\sla A_{2}\sla
k_{2} \sla k_{3}W(1)\nonumber\\
F:&&-\f{1}{2}k_{1}\cdot k_{2} W(4)\sla A_{3}\sla k_{3} \sla
A_{2}W(1) +\f{1}{2}A_{2}\cdot k_{1} W(4)\sla A_{3}\sla
k_{3} \sla k_{2}W(1)
\end{eqnarray}
Each of $C+D$ and $E+F$ can then be re-expressed as
\begin{eqnarray}
&&\f{1}{2}k_{1}\cdot k_{4}~W(1)\sla A_{3}\sla k_{3}\sla
A_{2}W(4)\nonumber\\
&+&k_{1}\cdot k_{4}~k_{4}\cdot A_{2}~ W(1)\sla
A_{3}W(4)+k_{1}\cdot k_{2}~A_{2}\cdot
A_{3}~ W(1)\sla k_{2}W(4)\nonumber\\
&+&k_{1}\cdot A_{2}~k_{4}\cdot A_{3}~ W(1)\sla
k_{2}W(4)+k_{1}\cdot A_{3}~k_{4}\cdot A_{2}~ W(1)\sla
k_{3}W(4)
\end{eqnarray}
Another expression for each of $C,D,E,F$ can be obtained by
absorbing the second term into the first term, and the summed
result is:
\begin{equation}
-k_{1}\cdot k_{3}~W(1)\sla A_{3}(\sla k_{3}+\sla k_{4})\sla
A_{2}W(4)-k_{1}\cdot k_{2} W(1)\sla A_{2}(\sla k_{2}+\sla
k_{4})\sla A_{3}W(4)
\end{equation}

\mysubsection{1 loop : 4 fermions}

There are totally
$\f12\cdot{4\choose 2}\cdot2\cdot2=3\cdot 4=12$ terms (and also 12
corresponding $(dF)^{2}W^{2}$ terms) contributing to the 4-fermion
amplitude:
\begin{eqnarray}
[\alpha\beta][\gamma\delta]&:&\f{1}{2!2!}(d_{[\delta}d_{\gamma]}W^{\alpha})W^{\beta}(d_{[\beta}d_{\alpha]}W^{\gamma})W^{\delta}\nonumber\\
&&-\f{1}{2!2!}(d_{[\gamma}d_{\delta]}W^{\alpha})W^{\beta}W^{\gamma}(d_{[\beta}d_{\alpha]}W^{\delta})\nonumber\\
&&-\f{1}{2!2!}W^{\alpha}(d_{[\delta}d_{\gamma]}W^{\beta})(d_{[\alpha}d_{\beta]}W^{\gamma})W^{\delta}\nonumber\\
&&+\f{1}{2!2!}W^{\alpha}(d_{[\gamma}d_{\delta]}W^{\beta})W^{\gamma}(d_{[\alpha}d_{\beta]}W^{\delta})\nonumber\\~
[\alpha\gamma][\beta\delta]&:&-\f{1}{2!2!}(d_{[\delta}d_{\beta]}W^{\alpha})(d_{[\gamma}d_{\alpha]}W^{\beta})W^{\gamma}W^{\delta}\nonumber\\
&&+\f{1}{2!2!}(d_{[\beta}d_{\delta]}W^{\alpha})W^{\beta}W^{\gamma}(d_{[\gamma}d_{\alpha]}W^{\delta})\nonumber\\
&&+\f{1}{2!2!}W^{\alpha}(d_{[\alpha}d_{\gamma]}W^{\beta})(d_{[\delta}d_{\beta]}W^{\gamma})W^{\delta}\nonumber\\
&&-\f{1}{2!2!}W^{\alpha}W^{\beta}(d_{[\beta}d_{\delta]}W^{\gamma})(d_{[\alpha}d_{\gamma]}W^{\delta})\nonumber\\~
[\alpha\delta][\beta\gamma]&:&-\f{1}{2!2!}W^{\alpha}(d_{[\alpha}d_{\delta]}W^{\beta})W^{\gamma}(d_{[\gamma}d_{\beta]}W^{\delta})\nonumber\\
&&+\f{1}{2!2!}(d_{[\gamma}d_{\beta]}W^{\alpha})(d_{[\delta}d_{\alpha]}W^{\beta})W^{\gamma}W^{\delta}\nonumber\\
&&-\f{1}{2!2!}(d_{[\beta}d_{\gamma]}W^{\alpha})W^{\beta}(d_{[\delta}d_{\alpha]}W^{\gamma})W^{\delta}\nonumber\\
&&+\f{1}{2!2!}W^{\alpha}W^{\beta}(d_{[\alpha}d_{\delta]}W^{\gamma})(d_{[\beta}d_{\gamma]}W^{\delta})
\end{eqnarray}
For each term there are 4 terms, which come from
$[d_{\alpha}d_{\beta}-\gamma^{a}_{\alpha\beta}(-i\partial_{a})]^{2}$.
Among them only the $\gamma\gamma$ term survives, and the others
vanish due to a GP sum from corresponding $(dF)^{2}WW$ terms.

For each group two terms are equal to the other 2 terms, and the
resultant 6 terms are
\begin{eqnarray}
\label{4fermistep}
&&2W(1)\gamma_{a}W(2)~W(3)\gamma_{b}W(4)~k^{b}_{1}k^{a}_{3}\nonumber\\
&&-2W(1)\gamma_{a}W(2)~W(3)\gamma_{b}W(4)~k^{b}_{1}k^{a}_{4}\nonumber\\
&&+2W(1)\gamma_{a}W(3)~W(2)\gamma_{b}W(4)~k^{b}_{1}k^{a}_{2}\nonumber\\
&&-2W(1)\gamma_{a}W(3)~W(2)\gamma_{b}W(4)~k^{b}_{1}k^{a}_{4}\nonumber\\
&&-2W(1)\gamma_{a}W(4)~W(2)\gamma_{b}W(3)~k^{b}_{1}k^{a}_{3}\nonumber\\
&&+2W(1)\gamma_{a}W(4)~W(2)\gamma_{b}W(3)~k^{b}_{1}k^{a}_{2}
\end{eqnarray}
Symmetry between any two fermion lines is somewhat obscure in this
form. But there are Fierz identities which make it clear:
\begin{eqnarray*}
A:&&W(1)\gamma^{a}W(2)~W(3)\gamma_{a}W(4)\nonumber\\
&=&-W(1)\gamma^{a}W(3)~W(4)\gamma_{a}W(2)-W(1)\gamma^{a}W(4)~W(2)\gamma_{a}W(3)\hspace{1.6cm}
\end{eqnarray*}
\begin{eqnarray*}
B:&&W(2)\gamma^{c}\gamma_{d}\gamma^{a}W(1)~W(4)\gamma_{a}\gamma_{b}\gamma_{c}W(3)k_{1}^{b}k^{d}_{3}\\
&=&4W(2)\gamma^{a}W(4)~W(3)\gamma_{a}W(1)k_{2}\cdot k_{3}
-4W(2)\gamma_{a}W(4)~W(3)\gamma_{b}W(1)k_{3}^{a}k^{b}_{2}\\
&&+12W(2)\gamma_{a}W(3)~W(1)\gamma W(4)k^{a}_{1}k^{b}_{3}
-12W(2)\gamma^{a}W_{3}~W(1)\gamma_{a}W(4)k_{1}\cdot k_{3}
\end{eqnarray*}
\begin{eqnarray}
C:&&W(2)\gamma^{c}\gamma_{d}\gamma^{a}W(1)~W(4)\gamma_{a}\gamma_{b}\gamma_{c}W(3)k_{1}^{b}k^{d}_{3}\nonumber\\
&=&8W(2)\gamma^{a}W(4)~W(3)\gamma_{a}W(1)k_{2}\cdot k_{3}
-8W(2)\gamma_{a}W(4)~W(3)\gamma_{b}W(1)k_{3}^{a}k^{b}_{2}\nonumber\\
&&+16W(2)\gamma_{a}W(3)~W(1)\gamma
W(4)k^{a}_{1}k^{b}_{3}-16W(2)\gamma^{a}W_{3}~W(1)\gamma_{a}W(4)k_{1}\cdot
k_{3}\nonumber\\
&&-4W(2)\gamma_{a}W(1)~W(4)\gamma_{b}W(3)k_{3}^{a}k_{1}^{b}
\end{eqnarray}
Using the above identities we can rewrite (\ref{4fermistep}) as
\begin{equation}
4k_{1}\cdot k_{2}~W(1)\gamma W(4)\cdot W(2)\gamma W(3)
-4k_{1}\cdot k_{4}~W(1)\gamma W(2)\cdot W(3)\gamma W(4)
\end{equation}
Now symmetry in fermion lines can be checked using Fierz identity
$A$.

\end{document}